%% file: alg.t-d.tex
\documentclass[journal]{IEEEtran}
\ifCLASSINFOpdf
\else
    \usepackage[dvips]{graphicx}
\fi
\newcommand{\bysame}{%
    \leavevmode\hbox to 3em{\hrulefill}\,}

\begin{document}
\title{Algebraic Construction of Tail-Biting Trellises for Linear Block Codes}

\author{Masato~Tajima,~\IEEEmembership{Senior~Member,~IEEE}
        %John~Doe,~\IEEEmembership{Fellow,~OSA,}
        %and~Jane~Doe,~\IEEEmembership{Life~Fellow,~IEEE}% <-this % stops a space
\thanks{M. Tajima was with the Graduate School of
Science and Engineering, University of Toyama, 3190 Gofuku,
Toyama 930-8555, Japan (e-mail: masatotjm@kind.ocn.ne.jp).}% <-this % stops a space
\thanks{}}

\maketitle

\begin{abstract}
In this paper, we present an algebraic construction of tail-biting trellises. The proposed method is based on the state space expressions, i.e., the state space is the image of the set of information sequences under the associated state matrix. Then combining with the homomorphism theorem, an algebraic trellis construction is obtained. We show that a tail-biting trellis constructed using the proposed method is isomorphic to the associated Koetter--Vardy (KV) trellis and tail-biting Bahl--Cocke--Jelinek--Raviv (BCJR) trellis. We also evaluate the complexity of the obtained tail-biting trellises. On the other hand, a matrix consisting of linearly independent rows of the characteristic matrix is regarded as a generalization of minimal-span generator matrices. Then we show that a KV trellis is constructed based on an extended minimal-span generator matrix. It is shown that this construction is a natural extension of the method proposed by McEliece (1996).
\end{abstract}


% \fi
\IEEEpeerreviewmaketitle

\section{Introduction}
\IEEEPARstart{F}{rom} the 1980s to 1990s, trellis representations of linear block codes were studied with a great interest~\cite{forn 88,forn 94,kie 96,ks 95,lin 98,mc 96,mud 88,zya 93}. Subsequently, tail-biting trellises of linear block codes have received much attention. There have been many contributions to the subject~\cite{con 15,glu 111,glu 112,glu 13,koe 03,lin 00,nori 06,sha 00,taji 152,wea 12}. Given a linear block code, there exists a unique minimal conventional trellis. This trellis simultaneously minimizes all measures of trellis complexity. However, tail-biting trellises do not have such a property. That is, minimality of tail-biting trellises depends on the measure being used~\cite{koe 03}. Despite these difficulties, tail-biting trellises have been studied with a great interest. This is due to the fact that the complexity of a tail-biting trellis may be much lower than that of the minimal conventional trellis. A remarkable advance has been made by Koetter and Vardy~\cite{koe 03}. They showed that for a $k$-dimensional linear block code of length $n$ with full support, there exists a list of $n$ characteristic generators (i.e., a {\it characteristic matrix}~\cite{koe 03}) from which all minimal tail-biting trellises can be obtained. A different method of producing tail-biting trellises was proposed by Nori and Shankar~\cite{nori 06}. They used the {\it Bahl-Cocke-Jelinek-Raviv} (BCJR) construction~\cite{bahl 74} in order to obtain tail-biting trellises. In this construction, each path in the conventional trellis is displaced using a {\it displacement matrix}~\cite{nori 06} which is defined based on the spans of a generator matrix of the given code. These works were further investigated by Gluesing-Luerssen and Weaver~\cite{glu 111,glu 112}. They carefully examined the works by Koetter and Vardy and by Nori and Shankar. In particular, they noted the fact that the characteristic matrix associated with a given code is not unique in general. Taking account of this fact, they have refined and generalized the previous works.
\par
This paper focuses on algebraic constructions of tail-biting trellises. In 1988, Forney~\cite{forn 88}, in an appendix to a paper on coset codes, provided an algebraic characterization of conventional trellises, which resulted in a great interest in the subject. In connection with algebraic trellis constructions, Nori and Shankar~\cite{nori 06} discussed a generalization of the Forney construction~\cite{forn 88} to tail-biting trellises. On the other hand, the state and edge spaces of a tail-biting trellis have been characterized by Gluesing-Luerssen and Weaver~\cite{glu 111,glu 112}. Let $M_i$ be the state matrix at level $i$ of a {\it Koetter-Vardy} (KV) trellis of a linear block code $C$. Then the state space is given by $V_i=\mbox{im}M_i$ (i.e., the image of $\mbox{F}^k$ under the linear mapping $M_i$, where $\mbox{F}^k$ denotes the set of information sequences of length $k$). Similarly, let $N_i$ be the state matrix at level $i$ of a tail-biting BCJR trellis of $C$. Then the state space is given by $V_i=\mbox{im}N_i$. From these expressions, we noticed that the {\it homomorphism theorem} can be applied to $V_i=\mbox{im}M_i~(V_i=\mbox{im}N_i)$. That is, we have
\begin{eqnarray}
V_i &=& \mbox{im}M_i\cong \mbox{F}^k/\mbox{ker}(M_i),~\mbox{for}~i=0, \cdots, n-1 \nonumber \\
V_i &=& \mbox{im}N_i\cong \mbox{F}^k/\mbox{ker}(N_i),~~\mbox{for}~i=0, \cdots, n-1, \nonumber
\end{eqnarray}
where $\mbox{ker}(M_i)$ and $\mbox{ker}(N_i)$ are the kernels of the linear mappings $M_i$ and $N_i$, respectively. These equations directly provide an algebraic construction of tail-biting trellises. In this paper, based on these fundamental relations, we propose an algebraic construction of tail-biting trellises for linear block codes. It is shown that a tail-biting trellis constructed using the proposed method is isomorphic to the associated KV trellis and tail-biting BCJR trellis. We also evaluate the complexity of the obtained tail-biting trellises.
\par
On the other hand, note that characteristic generators may be regarded as a generalization of {\it minimal-span generator matrices} (MSGM's) in the realm of conventional trellises~\cite{glu 112}. Hence, it is reasonable to think that a tail-biting trellis can also be constructed based on a kind of MSGM. Suppose that $G$ consists of $k$ linearly independent rows of a characteristic matrix. Then $G$ is regarded as a generalization of MSGM. We call such a generator matrix an {\it extended minimal-span generator matrix} (e-MSGM). We show that a KV trellis is constructed based on an e-MSGM. We also discuss the relationship between the proposed algebraic construction and the construction based on e-MSGM's.
\par
The paper is organized as follows. The basic notions for tail-biting trellises are given in Section II. In Section III, we first review the algebraic trellis construction by Nori and Shankar. After that we will present an algebraic construction of tail-biting trellises. The proposed construction is based on the state-space expressions combined with the homomorphism theorem. The complexity of tail-biting trellises obtained using the proposed construction will be evaluated in Section IV. In Section V, we discuss the tail-biting trellis construction based on e-MSGM's. Finally, conclusions are provided in Section VI.

\section{Preliminaries}
In this section, we introduce the basic notions needed in this paper. We always assume that the underlying field is $\mbox{F}=\mbox{GF}(2)$. Denote by $C$ an $(n, k)$ linear block code defined by a generator matrix $G$ and a corresponding parity-check matrix $H$. Let
\begin{eqnarray}
G &=& \left(
\begin{array}{c}
\mbox{\boldmath $g$}_1 \\
\mbox{\boldmath $g$}_2 \\
\cdots \\
\mbox{\boldmath $g$}_k
\end{array}
\right) \\
&=& \left(
\begin{array}{cccc}
\mbox{\boldmath $\bar g$}_1 & \mbox{\boldmath $\bar g$}_2 & \cdots & \mbox{\boldmath $\bar g$}_n
\end{array}
\right) \\
H &=& \left(
\begin{array}{cccc}
\mbox{\boldmath $h$}_1 & \mbox{\boldmath $h$}_2 & \cdots & \mbox{\boldmath $h$}_n
\end{array}
\right) .
\end{eqnarray}
A tail-biting trellis $T=(V, E)$ of depth $n$ over the field $\mbox{F}$ is a directed edge-labeled graph with the property that the vertex set $V$ partitions into $n$ disjoint sets $V=V_0 \cup V_1 \cup \cdots \cup V_{n-1}$. Here every edge in $T$ starts in $V_{i-1}$ and ends in $V_{i~mod~n}$. An edge is a triple $(v, a, w)\in V_{i-1}\times \mbox{F} \times V_i$. We call $V_i$ the {\it state space} of the trellis at level $i$. Thus its elements are called the {\it states} at that level. A {\it cycle} in $T$ is a closed path of length $n$. We assume that the cycles start and end at the same state in $V_0$. If $\vert V_0 \vert =1$, the trellis is called {\it conventional}.
\par
In addition to the labeling of edges, each vertex in $V_i$ can also be labeled. The resulting trellis is termed a labeled trellis. Then every cycle in a labeled tail-biting trellis $T$ consists of the labels of edges and vertices in the cycle. Such a sequence is termed a label sequence in $T$. The set of all the label sequences in a labeled tail-biting trellis $T$ is called the label code of $T$, denoted by $S(T)$. We call a trellis {\it reduced} if every state and every edge appears in at least one cycle. A labeled trellis $T$ is said to be {\it linear}, if $T$ is reduced and $S(T)$ is a linear code over $\mbox{F}$. Linear trellises $T=(V, E)$ and $T'=(V', E')$ are called {\it isomorphic} if there exists a bijection $\phi:~V \rightarrow V'$ such that (a restriction) $\phi_{\vert V_i}:~V_i\rightarrow V_i'~(0\leq i \leq n-1)$ is an isomorphism and $(v, a, w)\in E_i\leftrightarrow (\phi(v), a, \phi(w))\in E_i'$ for all $i~(1\leq i \leq n)$.
\par
Given a codeword $\mbox{\boldmath $x$}\in C$, a span of {\boldmath $x$}, denoted $[\mbox{\boldmath $x$}]$, is a semiopen interval $(a, b]$ such that the corresponding closed interval $[a, b]$ contains all the nonzero positions of {\boldmath $x$}. We call the intervals $(a, b]$ and $[a, b]$ {\it conventional} if $a\leq b$ and {\it circular} otherwise.
\par
{\it Remark 1:} Note that $[\mbox{\boldmath $x$}]$ does not contain the starting point $a$ in the span. This is very convenient for the definition of elementary trellises~\cite{koe 03}. On the other hand, a closed interval $[a, b]$ is adopted as a span in~\cite{mc 96} and~\cite{nori 06}. Hence, we use the latter, if necessary. Also, take notice of the numbering of indices for a codeword. We assume that the index starts in $0$ and ends in $n-1$ for KV trellises.
\par
Let $X~(\in \mbox{F}^{n\times n})$ be a {\it characteristic matrix}~\cite{koe 03} of $C$. The rows of $X$ are called characteristic generators. Let $(a_l, b_l]$ be the span of a characteristic generator $\mbox{\boldmath $g$}_l$. Note that $a_1, \cdots, a_n$ are distinct and $b_1, \cdots, b_n$ are distinct. Let $T_{g_l,(a_l, b_l]}$ be the {\it elementary trellis}~\cite{koe 03} corresponding to a characteristic generator $\mbox{\boldmath $g$}_l$. A trellis of the form $T_{g_{l_1},(a_{l_1}, b_{l_1}]}\times \cdots \times T_{g_{l_k},(a_{l_k}, b_{l_k}]}$, where $\mbox{\boldmath $g$}_{l_1}, \cdots, \mbox{\boldmath $g$}_{l_k}$ are linearly independent rows of $X$, is called a {\it KV trellis}~\cite{glu 111,koe 03} of $C$.
\par
{\it Remark 2:} The name KV trellises is used more generally for product trellises of the type $T_{G, S}$, where $G$ is a generator matrix and $S$ is the corresponding span list.
\par
Next, we consider a tail-biting BCJR trellis introduced by Nori and Shankar~\cite{nori 06}. Denote by $G$ and $H$ the generator matrix and parity-check matrix of $C$, respectively. Let $S=\{[a_l, b_l],~1\leq l \leq k\}$ be a span list of $G$. A displacement matrix $\Theta$, defined by Nori and Shankar, is a design parameter for the construction of good trellises. In this paper, $\Theta$ is defined based on $S$ as follows (see~\cite{glu 111} or~\cite{nori 06}):
\begin{equation}
\Theta=\left(
\begin{array}{c}
\mbox{\boldmath $d$}_{g_1} \\
\mbox{\boldmath $d$}_{g_2} \\
\cdots \\
\mbox{\boldmath $d$}_{g_k} 
\end{array}
\right)\in \mbox{F}^{k\times (n-k)}~\mbox{with}~\mbox{\boldmath $d$}_l=\sum_{j=a_l}^ng_{lj}\mbox{\boldmath $h$}_j^T ,
\end{equation}
where $\mbox{\boldmath $g$}_l=(g_{l1}, \cdots, g_{ln})$ is a generator $\in G$ with span $[a_l, b_l]$ ($T$ means transpose). Note that if $\mbox{\boldmath $g$}_l$ has a conventional span, then $\mbox{\boldmath $d$}_l=\mbox{\boldmath $0$}$. The displacement vector $\mbox{\boldmath $d$}_c$ for any codeword $\mbox{\boldmath $c$} \in C$ is defined as follows~\cite{nori 06}:
\begin{equation}
\mbox{\boldmath $d$}_c=\sum_{i=1}^k\alpha_i\mbox{\boldmath $d$}_{g_i}, ~\mbox{where}~\mbox{\boldmath $c$}=\sum_{i=1}^k\alpha_i\mbox{\boldmath $g$}_i,~\alpha_i \in \mbox{F},~\mbox{\boldmath $g$}_i\in G .
\end{equation}
Denote by $T_{(G, H, \Theta)}$ the resulting trellis. The trellis $T_{(G, H, \Theta)}$ is called a tail-biting BCJR trellis~\cite{glu 111,nori 06}.

\input{sec.3-d}
\input{sec.4-d}
\input{sec.5-c}

\section{Conclusion}
We have presented an algebraic construction of tail-biting trellises. The proposed method is based on a quite simple idea. We took notice of the state space expressions of a tail-biting trellis, where the state space is the image of the set of information sequences under the associated state matrix. Then by applying the homomorphism theorem to these expressions, an algebraic trellis construction is obtained. We have shown that a tail-biting trellis constructed using the proposed method is isomorphic to the associated KV trellis and tail-biting BCJR trellis. Also, we have evaluated the complexity of the obtained tail-biting trellises. On the other hand, a matrix consisting of linearly independent rows of the characteristic matrix is regarded as a generalization of minimal-span generator matrices. Then we have shown that a KV trellis is constructed based on an extended minimal-span generator matrix. It is shown that this construction is a natural extension of the method proposed by McEliece~\cite[Section VII]{mc 96}.

% or


\appendices
\section{Proof of Lemma 2}
\input{sec.app}

%\section{}

\section*{Acknowledgment}
The author would like to thank Prof. H. Gluesing-Luerssen for many useful comments.

\ifCLASSOPTIONcaptionsoff
  \newpage
\fi

\end{document}

%% file: sec.3-d.tex
\section{Algebraic Construction of Tail-Biting Trellises}
Nori and Shankar~\cite[Section V]{nori 06} generalized the Forney construction~\cite{forn 88} for conventional trellises to tail-biting trellises. In this section, we will attempt to do the same thing in a somewhat different way.

\subsection{Review of the tail-biting Forney trellis introduced by Nori and Shankar}
Consider the $(7, 4)$ Hamming code $C$ defined by the following generator matrix $G$ and parity-check matrix $H$ (cf.~\cite[Examples 14 and 16]{nori 06}):
\begin{eqnarray}
H &=& \left(
\begin{array}{ccccccc}
1& 1 & 0 & 0 & 1 & 0 & 1 \\
1& 1 & 1 & 0 & 0 & 1 & 0 \\
0& 1 & 1 & 1 & 0 & 0 & 1
\end{array}
\right) \\
&=& \left(
\begin{array}{ccccccc}
\mbox{\boldmath $h$}_1 & \mbox{\boldmath $h$}_2 & \mbox{\boldmath $h$}_3 & \mbox{\boldmath $h$}_4 & \mbox{\boldmath $h$}_5 & \mbox{\boldmath $h$}_6 & \mbox{\boldmath $h$}_7
\end{array}
\right) \nonumber
\end{eqnarray}
\begin{eqnarray}
G &=& \left(
\begin{array}{ccccccc}
\mbox{\boldmath $1$}& \mbox{\boldmath $0$} & \mbox{\boldmath $0$} & \mbox{\boldmath $0$} & \mbox{\boldmath $1$} & \mbox{\boldmath $1$} & 0 \\
0& 0 & \mbox{\boldmath $1$} & \mbox{\boldmath $0$} & \mbox{\boldmath $1$} & \mbox{\boldmath $1$} & \mbox{\boldmath $1$} \\
\mbox{\boldmath $0$}& \mbox{\boldmath $1$} & 0 & 0 & 0 & \mbox{\boldmath $1$} & \mbox{\boldmath $1$} \\
\mbox{\boldmath $0$}& \mbox{\boldmath $1$} & \mbox{\boldmath $1$} & \mbox{\boldmath $1$} & 0 & 0 & \mbox{\boldmath $1$}
\end{array}
\right)
\begin{array}{l}
$[1, 6]$ \\
$[3, 7]$ \\
$[6, 2]$ \\
$[7, 4]$
\end{array} \\
&=& \left(
\begin{array}{c}
\mbox{\boldmath $g$}_1 \\
\mbox{\boldmath $g$}_2 \\
\mbox{\boldmath $g$}_3 \\
\mbox{\boldmath $g$}_4
\end{array}
\right) . \nonumber
\end{eqnarray}
First, we construct the tail-biting BCJR trellis introduced by Nori and Shankar~\cite{nori 06}. Note that $\mbox{\boldmath $g$}_1$ and $\mbox{\boldmath $g$}_2$ have conventional spans, whereas $\mbox{\boldmath $g$}_3$ and $\mbox{\boldmath $g$}_4$ have circular spans $[6, 2]$ and $[7, 4]$, respectively. Hence, the corresponding displacement matrix $\Theta$ is given by
\begin{eqnarray}
\Theta &=& \left(
\begin{array}{c}
\mbox{\boldmath $d$}_{g_1} \\
\mbox{\boldmath $d$}_{g_2} \\
\mbox{\boldmath $d$}_{g_3} \\
\mbox{\boldmath $d$}_{g_4} 
\end{array}
\right) \nonumber \\
&=& \left(
\begin{array}{ccc}
0& 0 & 0 \\
0& 0 & 0 \\
1& 1 & 1 \\
1& 0 & 1 
\end{array}
\right) .
\end{eqnarray}
The resulting tail-biting BCJR trellis $T_{(G, H, \Theta)}$ is shown in Fig.1.
\begin{figure}[tb]
\begin{center}
\includegraphics[width=7.0cm,clip]{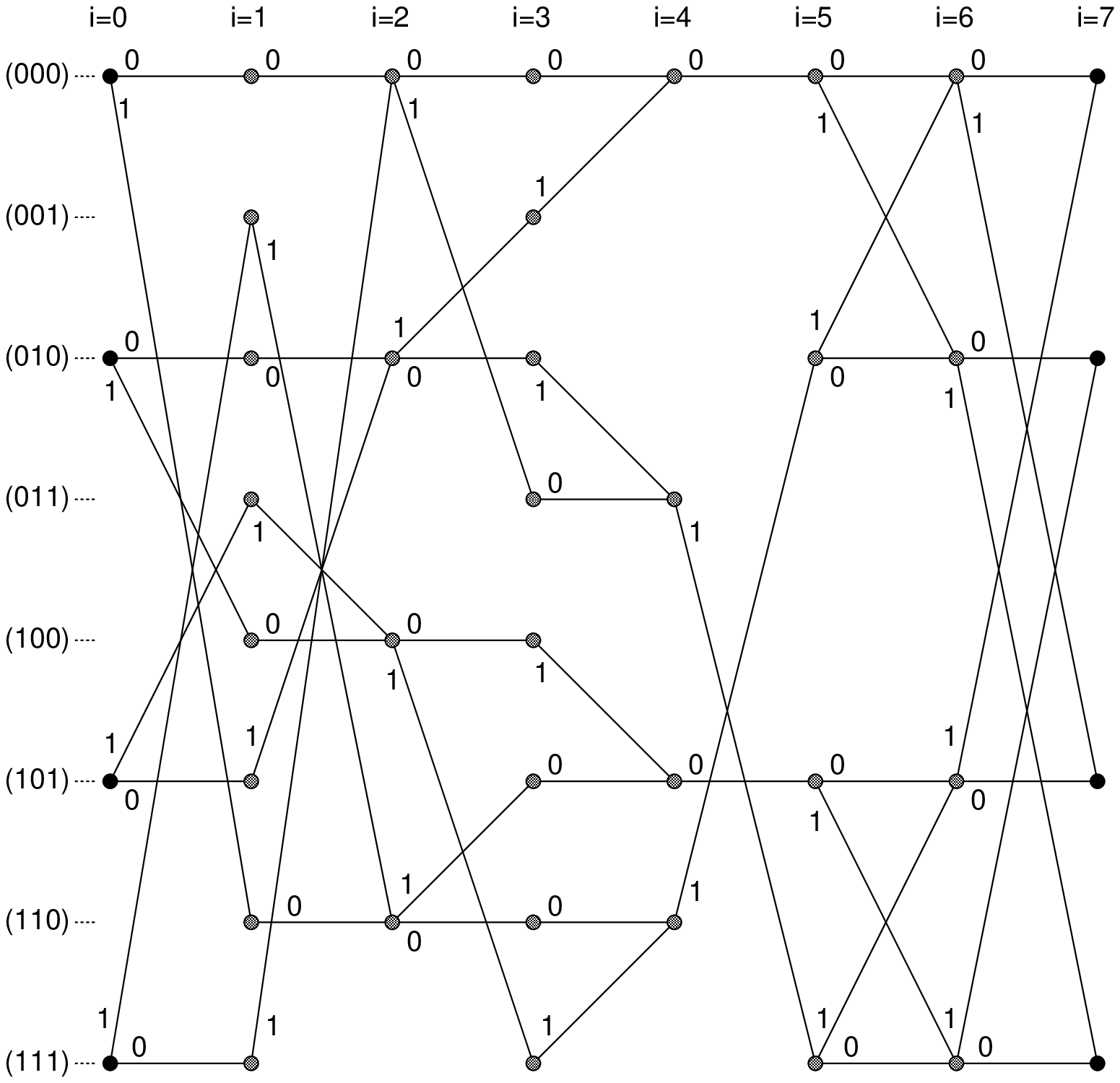}
\end{center}
\caption{The tail-biting BCJR trellis based on $(G, H, \Theta)$.}
\label{fig:1}
\end{figure}
\par
On the other hand, $G$ can be decomposed as follows:
\begin{equation}
G=\left(
\begin{array}{c}
G_0 \\
G_d
\end{array}
\right) ,
\end{equation}
where
\begin{eqnarray}
G_0 &=& \left(
\begin{array}{c}
\mbox{\boldmath $g$}_1 \\
\mbox{\boldmath $g$}_2
\end{array}
\right) \\
G_d &=& \left(
\begin{array}{c}
\mbox{\boldmath $g$}_3 \\
\mbox{\boldmath $g$}_4
\end{array}
\right) .
\end{eqnarray}
Here let $C_0$ be the linear subcode generated by $G_0$ and then consider the partition $C/C_0=\{C_0, C_1, C_2, C_3\}$. We have
\begin{eqnarray}
C_0 &=& \{0000000, 0010111, 1000110, 1010001\} \nonumber \\
C_1 &=& \{0100011, 0110100, 1100101, 1110010\} \nonumber \\
C_2 &=& \{0111001, 0101110, 1111111, 1101000\} \nonumber \\
C_3 &=& \{0011010, 0001101, 1011100, 1001011\} . \nonumber
\end{eqnarray}
\par
{\it Remark 1:} In this paper, we regard the first element of each coset as the representative. 
\par
We see that $G_d$ is a generator matrix for the set of representatives of the cosets $C_l~(0\leq l \leq 3)$.
\par
Next, for the cosets $C_l,~0\leq l \leq 3$, we define the following mappings:
\begin{eqnarray}
\pi_{i, 0}(\mbox{\boldmath $c$}) &=& c_1\mbox{\boldmath $h$}_1^T+\cdots +c_i\mbox{\boldmath $h$}_i^T \\
\pi_{i, 1}(\mbox{\boldmath $c$}) &=& c_1\mbox{\boldmath $h$}_1^T+\cdots +c_i\mbox{\boldmath $h$}_i^T+\mbox{\boldmath $d$}_{g_3} \\
\pi_{i, 2}(\mbox{\boldmath $c$}) &=& c_1\mbox{\boldmath $h$}_1^T+\cdots +c_i\mbox{\boldmath $h$}_i^T+\mbox{\boldmath $d$}_{g_4} \\
\pi_{i, 3}(\mbox{\boldmath $c$}) &=& c_1\mbox{\boldmath $h$}_1^T+\cdots +c_i\mbox{\boldmath $h$}_i^T+\mbox{\boldmath $d$}_{g_3}+\mbox{\boldmath $d$}_{g_4} ,
\end{eqnarray}
where $\mbox{\boldmath $c$}=(c_1, c_2, \cdots, c_7)$ denotes any codeword in coset $C_l$.
\par
For each $i$, $\bar C_{i0}=\bigcup_{0\leq l \leq 3}\{\mbox{\boldmath $c$}\in C_l:\pi_{i, l}(\mbox{\boldmath $c$})=\mbox{\boldmath $0$}\}$ is computed as follows:
\begin{eqnarray}
i=0 : \bar C_{00} &=& \{0000000, 0010111, 1000110, 1010001\} \nonumber \\
i=1 : \bar C_{10} &=& \{0000000, 0010111\} \nonumber \\
i=2 : \bar C_{20} &=& \{0000000, 0010111, 0100011, 0110100\} \nonumber \\
i=3 : \bar C_{30} &=& \{0000000, 0100011\} \nonumber \\
i=4 : \bar C_{40} &=& \{0000000, 0100011, 0111001, 0011010\} \nonumber \\
i=5 : \bar C_{50} &=& \{0000000, 0100011, 0111001, 0011010\} \nonumber \\
i=6 : \bar C_{60} &=& \{0000000, 1000110, 0111001, 1111111\}. \nonumber 
\end{eqnarray}
Note that $\bar C_{i0}$ is a linear subcode of $C$ for each $i$. Hence, we can think of the partition $C/\bar C_{i0}$. We have the following:
\begin{eqnarray}
i=0 : \bar C_{00} &=& \{0000000, 0010111, 1000110, 1010001\} \nonumber \\
\bar C_{01} &=& \{0111001, 0101110, 1111111, 1101000\} \nonumber \\
\bar C_{02} &=& \{0100011, 0110100, 1100101, 1110010\} \nonumber \\
\bar C_{03} &=& \{0011010, 0001101, 1011100, 1001011\} \nonumber \\
i=1 : \bar C_{10} &=& \{0000000, 0010111\} \nonumber \\
\bar C_{11} &=& \{1000110, 1010001\} \nonumber \\
\bar C_{12} &=& \{0111001, 0101110\} \nonumber \\
\bar C_{13} &=& \{1111111, 1101000\} \nonumber \\
\bar C_{14} &=& \{0100011, 0110100\} \nonumber \\
\bar C_{15} &=& \{1100101, 1110010\} \nonumber \\
\bar C_{16} &=& \{0011010, 0001101\} \nonumber \\
\bar C_{17} &=& \{1011100, 1001011\} \nonumber \\
i=2 : \bar C_{20} &=& \{0000000, 0100011, 0010111, 0110100\} \nonumber \\
\bar C_{21} &=& \{0111001, 0011010, 0101110, 0001101\} \nonumber \\
\bar C_{22} &=& \{1000110, 1100101, 1010001, 1110010\} \nonumber \\
\bar C_{23} &=& \{1111111, 1011100, 1101000, 1001011\} \nonumber \\
i=3 : \bar C_{30} &=& \{0000000, 0100011\} \nonumber \\
\bar C_{31} &=& \{0111001, 0011010\} \nonumber \\
\bar C_{32} &=& \{0010111, 0110100\} \nonumber \\
\bar C_{33} &=& \{0101110, 0001101\} \nonumber \\
\bar C_{34} &=& \{1000110, 1100101\} \nonumber \\
\bar C_{35} &=& \{1111111, 1011100\} \nonumber \\
\bar C_{36} &=& \{1010001, 1110010\} \nonumber \\
\bar C_{37} &=& \{1101000, 1001011\} \nonumber \\
i=4 : \bar C_{40} &=& \{0000000, 0111001, 0100011, 0011010\} \nonumber \\
\bar C_{41} &=& \{0010111, 0101110, 0110100, 0001101\} \nonumber \\
\bar C_{42} &=& \{1000110, 1111111, 1100101, 1011100\} \nonumber \\
\bar C_{43} &=& \{1010001, 1101000, 1110010, 1001011\} \nonumber \\
i=5 : \bar C_{50} &=& \{0000000, 0111001, 0100011, 0011010\} \nonumber \\
\bar C_{51} &=& \{0010111, 0101110, 0110100, 0001101\} \nonumber \\
\bar C_{52} &=& \{1000110, 1111111, 1100101, 1011100\} \nonumber \\
\bar C_{53} &=& \{1010001, 1101000, 1110010, 1001011\} \nonumber \\
i=6 : \bar C_{60} &=& \{0000000, 0111001, 1000110, 1111111\} \nonumber \\
\bar C_{61} &=& \{0100011, 0011010, 1100101, 1011100\} \nonumber \\
\bar C_{62} &=& \{0010111, 0101110, 1010001, 1101000\} \nonumber \\
\bar C_{63} &=& \{0110100, 0001101, 1110010, 1001011\} . \nonumber 
\end{eqnarray}
\par
Now a tail-biting trellis can be constructed based on the above results. That is, for each codeword $\mbox{\boldmath $c$}\in C$, the corresponding path (i.e., cycle) is obtained by tracing the cosets $\bar C_{il}$ which contain {\boldmath $c$}. The obtained tail-biting trellis is shown in Fig.2. The states in Fig.2 are labeled by the representatives in the cosets $\bar C_{il},~0\leq i \leq 6$.

Also, a modified tail-biting trellis obtained using state permutations is shown in Fig.3. We observe that the trellis in Fig.3 is identical to the one in Fig.1.
\begin{figure}[htb]
\begin{center}
\includegraphics[width=7.5cm,clip]{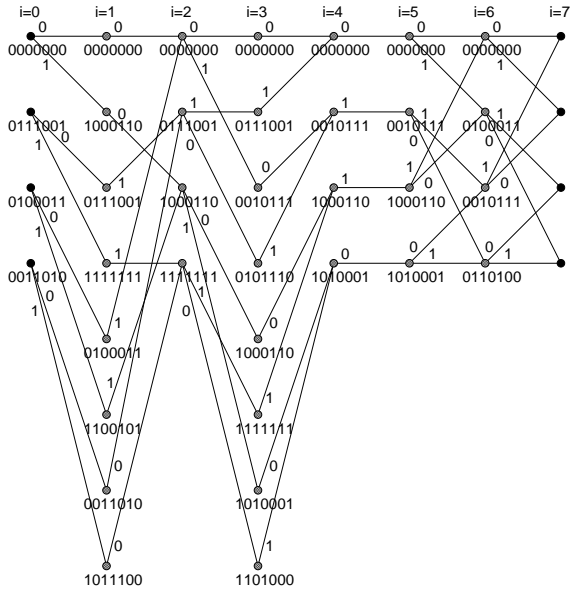}
\end{center}
\caption{The tail-biting trellis obtained using an algebraic construction.}
\label{fig:2}
\end{figure}
\begin{figure}[htb]
\begin{center}
\includegraphics[width=7.5cm,clip]{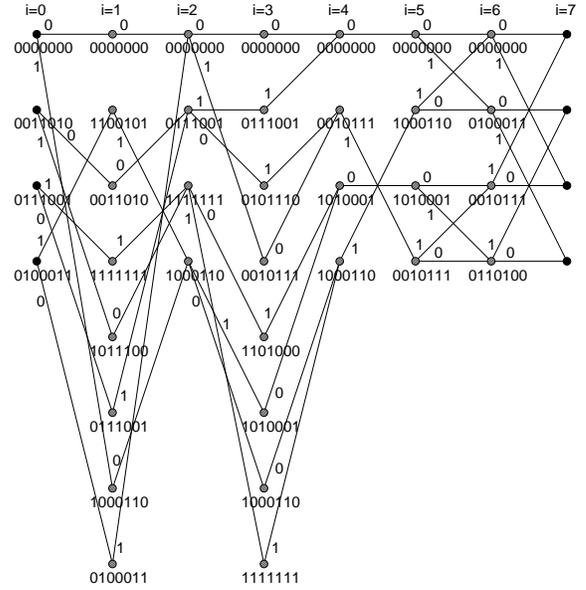}
\end{center}
\caption{A modified tail-biting trellis obtained using state permutations.}
\label{fig:3}
\end{figure}

\subsection{Algebraic construction of a tail-biting BCJR trellis}
The argument in the previous section, though it was presented in terms of a specific example, implies that a tail-biting BCJR trellis can be constructed using an algebraic method. Let $C$ be an $(n, k)$ linear block code defined by the generator matrix $G$ and parity-check matrix $H$. Denote by $\Theta$ the associated displacement matrix determined by $(G, H)$ and by the span list of $G$. Consider the tail-biting BCJR trellis $T_{(G, H, \Theta)}$ constructed based on $(G, H, \Theta)$. In the following, let $A_i$ denote the submatrix consisting of the first $i$ columns of $A$. The state matrix at level $i$, denoted by $N_i$, is given by
\begin{equation}
N_i=G_iH_i^T+\Theta,~\mbox{for}~i=0, \cdots, n-1.
\end{equation}
In this case~\cite{glu 111,glu 112}, the state space at level $i$ is expressed as
\begin{equation}
V_i=\mbox{im}N_i,~\mbox{for}~i=0, \cdots, n-1.
\end{equation}
Using the homomorphism theorem, the right-hand side becomes
\begin{equation}
\mbox{im}N_i\cong \mbox{F}^k/\mbox{ker}(N_i),~\mbox{for}~i=0, \cdots, n-1,
\end{equation}
where $\mbox{ker}(N_i)$ is the kernel of the linear mapping induced by $N_i$. Then we have
\begin{equation}
V_i=\mbox{im}N_i\cong \mbox{F}^k/\mbox{ker}(N_i),~~\mbox{for}~i=0, \cdots, n-1.
\end{equation}
\par
{\it Remark 2:} Consider the example in the previous section. The kernels $\mbox{ker}(N_i)~(0\leq i \leq 6)$ are obtained as follows:
\begin{eqnarray}
\mbox{ker}(N_0) &=& \{0000, 0100, 1000, 1100\} \nonumber \\
\mbox{ker}(N_1) &=& \{0000, 0100\} \nonumber \\
\mbox{ker}(N_2) &=& \{0000, 0010, 0100, 0110\} \nonumber \\
\mbox{ker}(N_3) &=& \{0000, 0010\} \nonumber \\
\mbox{ker}(N_4) &=& \{0000, 0001, 0010, 0011\} \nonumber \\
\mbox{ker}(N_5) &=& \{0000, 0001, 0010, 0011\} \nonumber \\
\mbox{ker}(N_6) &=& \{0000, 0001, 1000, 1001\} . \nonumber
\end{eqnarray}
Though the elements of $\mbox{ker}(N_i)$ are $\mbox{\boldmath $u$}\mbox{'s} \in \mbox{F}^4$, these elements can be identified with the corresponding codewords $\mbox{\boldmath $c$}\mbox{'s} \in C$. We see that the resulting set of codewords coincides with $\bar C_{i0},~0\leq i \leq 6$. In other words, $\bar C_{i0}=\bigcup_{0\leq l \leq 3}\{\mbox{\boldmath $c$}\in C_l:\pi_{i, l}(\mbox{\boldmath $c$})=\mbox{\boldmath $0$}\}$ is equal to $\mbox{ker}(N_i)$.
\par
For $i=0, \cdots, n-1$, let
\begin{equation}
\mbox{F}^k/\mbox{ker}(N_i)=\{\bar C_{i0}, \bar C_{i1}, \cdots, \bar C_{i,m(i)-1}\} .
\end{equation}
Here it is assumed that the elements in $\bar C_{il}$ have been transformed into the corresponding codewords. In this case, a tail-biting trellis is constructed by tracing the cosets $\bar C_{il}$ which contain {\boldmath $c$} for each codeword $\mbox{\boldmath $c$}\in C$. More precisely, there is an edge $e\in E_i$ labeled $c_i$ from a vertex $v \in V_{i-1}$ to a vertex $w \in V_i$, if and only if there exists a codeword $\mbox{\boldmath $c$}=(c_1, \cdots, c_n) \in C$ such that $\mbox{\boldmath $c$} \in v \cap w$.
\par
Similarly, we have the following for the edge spaces $E_i$:
\begin{eqnarray}
E_i &=& \mbox{im}(N_{i-1}, \mbox{\boldmath $\bar g$}_i, N_i) \nonumber \\
&\cong& \mbox{F}^k/\mbox{ker}(N_{i-1}, \mbox{\boldmath $\bar g$}_i, N_i),~\mbox{for}~i=1, \cdots, n ,
\end{eqnarray}
where $\mbox{\boldmath $\bar g$}_i$ denotes the $i$th column of $G$.
\par
For the obtained tail-biting trellis, we have the following.
\newtheorem{pro}{Proposition}
\begin{pro}
A tail-biting trellis obtained using the proposed construction, denoted by $T_{alg}$, is isomorphic to the associated tail-biting BCJR trellis $T_{(G, H, \Theta)}$.
\end{pro}
\begin{IEEEproof}
The proposed method is based on the isomorphism:
\begin{displaymath}
\mbox{im}N_i\cong \mbox{F}^k/\mbox{ker}(N_i),~\mbox{for}~n=0, \cdots, n-1.
\end{displaymath}
For $\mbox{\boldmath $u$},~\mbox{\boldmath $u$}' \in \mbox{F}^k$, let
\begin{eqnarray}
\mbox{\boldmath $c$} &=& (c_1, \cdots, c_i, \cdots, c_n) \nonumber \\
\mbox{\boldmath $c$}' &=& (c_1', \cdots, c_i', \cdots, c_n') \nonumber
\end{eqnarray}
be the corresponding codewords. Here suppose that $\mbox{\boldmath $u$}-\mbox{\boldmath $u$}' \in \mbox{ker}(N_i)$. This means that $\mbox{\boldmath $c$}$ and $\mbox{\boldmath $c$}'$ are contained in the same coset, i.e., go through the same state at level $i$. On the other hand, $\mbox{\boldmath $u$}-\mbox{\boldmath $u$}' \in \mbox{ker}(N_i)$ is equivalent to $\mbox{\boldmath $u$}N_i=\mbox{\boldmath $u$}'N_i$. Hence, noting $N_i=G_iH_i^T+\Theta$, we have
\begin{eqnarray}
\lefteqn{c_1\mbox{\boldmath $h$}_1^T+\cdots +c_i\mbox{\boldmath $h$}_i^T+(u_1\mbox{\boldmath $d$}_{g_1}+\cdots+u_k\mbox{\boldmath $d$}_{g_k})} \nonumber \\
&& =c_1'\mbox{\boldmath $h$}_1^T+\cdots +c_i'\mbox{\boldmath $h$}_i^T+(u_1'\mbox{\boldmath $d$}_{g_1}+\cdots+u_k'\mbox{\boldmath $d$}_{g_k}) . \nonumber
\end{eqnarray}
The last equation means that $\mbox{\boldmath $c$}$ and $\mbox{\boldmath $c$}'$ define the same state at level $i$ in the tail-biting BCJR trellis. 
\end{IEEEproof}
\par
{\it Example 1 (Nori and Shankar~\cite{nori 06}):} Consider the $(7, 4)$ Hamming code defined by the following generator matrix $G$ and parity-check matrix $H$:
\begin{eqnarray}
H &=& \left(
\begin{array}{ccccccc}
1& 1 & 0 & 0 & 1 & 0 & 1 \\
1& 1 & 1 & 0 & 0 & 1 & 0 \\
0& 1 & 1 & 1 & 0 & 0 & 1
\end{array}
\right) \\
G &=& \left(
\begin{array}{ccccccc}
0& 0 & 0 & \mbox{\boldmath $1$} & \mbox{\boldmath $1$} & \mbox{\boldmath $0$} & \mbox{\boldmath $1$} \\
\mbox{\boldmath $1$}& \mbox{\boldmath $1$} & \mbox{\boldmath $0$} & \mbox{\boldmath $1$} & 0 & 0 & 0 \\
0& 0 & \mbox{\boldmath $1$} & \mbox{\boldmath $1$} & \mbox{\boldmath $0$} & \mbox{\boldmath $1$} & 0 \\
\mbox{\boldmath $1$}& \mbox{\boldmath $0$} & \mbox{\boldmath $1$} & 0 & 0 & 0 & \mbox{\boldmath $1$}
\end{array}
\right)
\begin{array}{l}
$[4, 7]$ \\
$[1, 4]$ \\
$[3, 6]$ \\
$[7, 3]$
\end{array} \\
&=& \left(
\begin{array}{c}
\mbox{\boldmath $g$}_1 \\
\mbox{\boldmath $g$}_2 \\
\mbox{\boldmath $g$}_3 \\
\mbox{\boldmath $g$}_4
\end{array}
\right) . \nonumber
\end{eqnarray}
Since $\mbox{\boldmath $g$}_4$ has a circular span $[7, 3]$, the corresponding displacement matrix is given by
\begin{eqnarray}
\Theta &=& \left(
\begin{array}{c}
\mbox{\boldmath $d$}_{g_1} \\
\mbox{\boldmath $d$}_{g_2} \\
\mbox{\boldmath $d$}_{g_3} \\
\mbox{\boldmath $d$}_{g_4} 
\end{array}
\right) \nonumber \\
&=& \left(
\begin{array}{ccc}
0& 0 & 0 \\
0& 0 & 0 \\
0& 0 & 0 \\
1& 0 & 1 
\end{array}
\right) .
\end{eqnarray}
The tail-biting BCJR trellis $T_{(G, H, \Theta)}$ is shown in Fig.4.
\begin{figure}[tb]
\begin{center}
\includegraphics[width=7.0cm,clip]{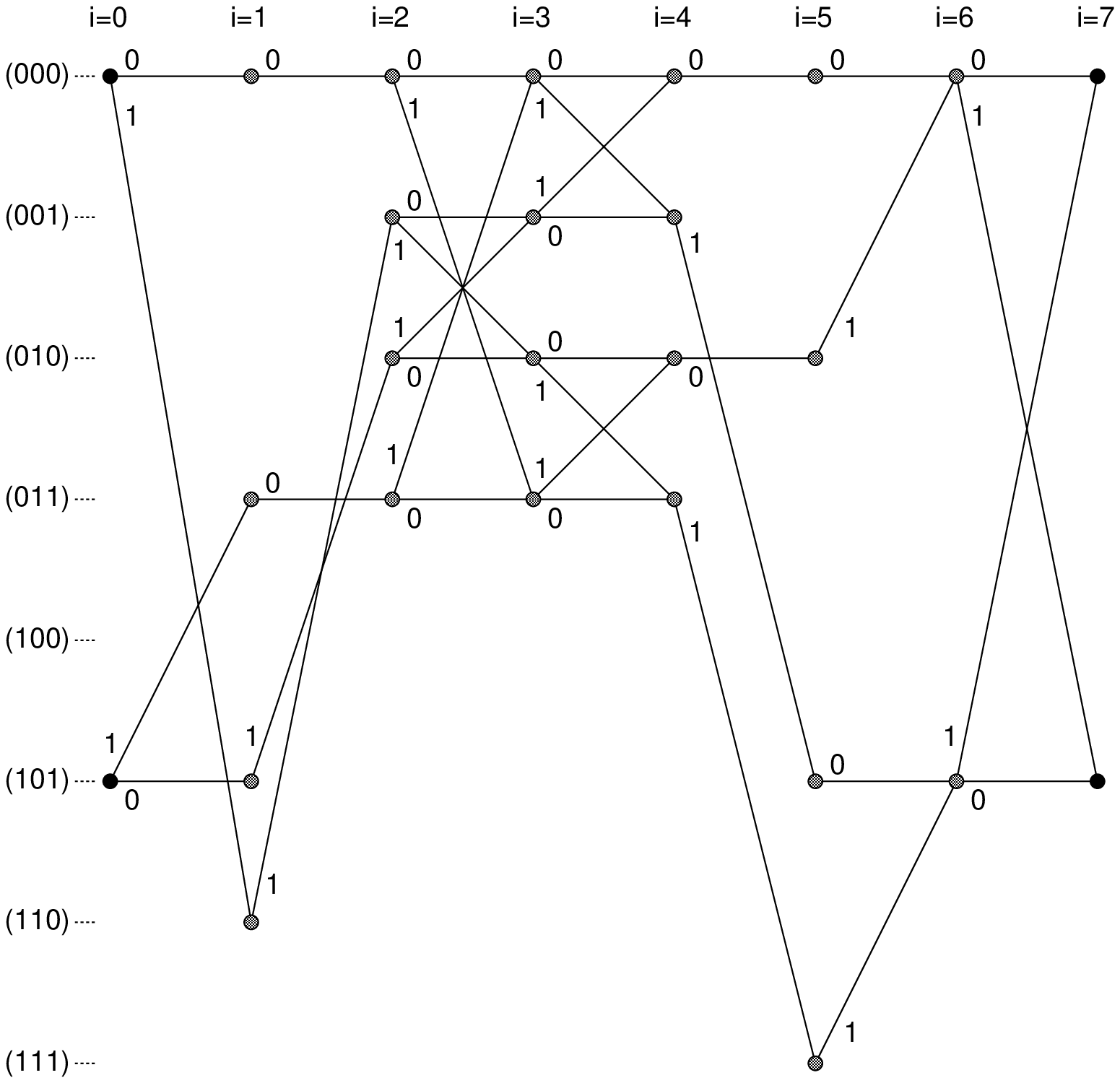}
\end{center}
\caption{The tail-biting BCJR trellis based on $(G, H, \Theta)$.}
\label{fig:4}
\end{figure}
\par
In order to construct a tail-biting trellis using the proposed method, we first compute the state matrices $N_i~(0\leq i \leq 6)$. They are given as follows:
\begin{eqnarray}
N_0 &=& \left(
\begin{array}{ccc}
0& 0 & 0 \\
0& 0 & 0 \\
0& 0 & 0 \\
1& 0 & 1 
\end{array}
\right)(=\Theta) \nonumber \\
N_1 &=& \left(
\begin{array}{ccc}
0& 0 & 0 \\
1& 1 & 0 \\
0& 0 & 0 \\
0& 1 & 1 
\end{array}
\right) \nonumber \\
N_2 &=& \left(
\begin{array}{ccc}
0& 0 & 0 \\
0& 0 & 1 \\
0& 0 & 0 \\
0& 1 & 1 
\end{array}
\right) \nonumber \\
N_3 &=& \left(
\begin{array}{ccc}
0& 0 & 0 \\
0& 0 & 1 \\
0& 1 & 1 \\
0& 0 & 0 
\end{array}
\right) \nonumber \\
N_4 &=& \left(
\begin{array}{ccc}
0& 0 & 1 \\
0& 0 & 0 \\
0& 1 & 0 \\
0& 0 & 0 
\end{array}
\right) \nonumber \\
N_5 &=& \left(
\begin{array}{ccc}
1& 0 & 1 \\
0& 0 & 0 \\
0& 1 & 0 \\
0& 0 & 0 
\end{array}
\right) \nonumber \\
N_6 &=& \left(
\begin{array}{ccc}
1& 0 & 1 \\
0& 0 & 0 \\
0& 0 & 0 \\
0& 0 & 0 
\end{array}
\right) . \nonumber
\end{eqnarray}
Next, $\mbox{ker}(N_i)~(0\leq i \leq 6)$ are determined as follows:
\begin{eqnarray}
\mbox{ker}(N_0) &=& \{0000, 0010, 0100, 0110, \nonumber \\
&& 1000, 1010, 1100, 1110\} \nonumber \\
\mbox{ker}(N_1) &=& \{0000, 0010, 1000, 1010\} \nonumber \\
\mbox{ker}(N_2) &=& \{0000, 0010, 1000, 1010\} \nonumber \\
\mbox{ker}(N_3) &=& \{0000, 0001, 1000, 1001\} \nonumber \\
\mbox{ker}(N_4) &=& \{0000, 0001, 0100, 0101\} \nonumber \\
\mbox{ker}(N_5) &=& \{0000, 0001, 0100, 0101\} \nonumber \\
\mbox{ker}(N_6) &=& \{0000, 0001, 0010, 0011, \nonumber \\
&& 0100, 0101, 0110, 0111\} . \nonumber
\end{eqnarray}
These kernels can also be expressed in terms of the corresponding codewords:
\begin{eqnarray}
i=0 : \bar C_{00} &=& \{0000000, 0011010, 1101000, 1110010, \nonumber \\
&& 0001101, 0010111, 1100101, 1111111\} \nonumber \\
i=1 : \bar C_{10} &=& \{0000000, 0011010, 0001101, 0010111\} \nonumber \\
i=2 : \bar C_{20} &=& \{0000000, 0011010, 0001101, 0010111\} \nonumber \\
i=3 : \bar C_{30} &=& \{0000000, 1010001, 0001101, 1011100\} \nonumber \\
i=4 : \bar C_{40} &=& \{0000000, 1010001, 1101000, 0111001\} \nonumber \\
i=5 : \bar C_{50} &=& \{0000000, 1010001, 1101000, 0111001\} \nonumber \\
i=6 : \bar C_{60} &=& \{0000000, 1010001, 0011010, 1001011, \nonumber \\
&& 1101000, 0111001, 1110010, 0100011\} . \nonumber
\end{eqnarray}
Then the partitions $C/\bar C_{i0}$ are given as follows:
\begin{eqnarray}
i=0 : \bar C_{00} &=& \{0000000, 0011010, 1101000, 1110010, \nonumber \\
&& 0001101, 0010111, 1100101, 1111111\} \nonumber \\
\bar C_{01} &=& \{1010001, 1001011, 0111001, 0100011, \nonumber \\
&& 1011100, 1000110, 0110100, 0101110\} \nonumber \\
i=1 : \bar C_{10} &=& \{0000000, 0011010, 0001101, 0010111\} \nonumber \\
\bar C_{11} &=& \{1010001, 1001011, 1011100, 1000110\} \nonumber \\
\bar C_{12} &=& \{1101000, 1110010, 1100101, 1111111\} \nonumber \\
\bar C_{13} &=& \{0111001, 0100011, 0110100, 0101110\} \nonumber \\
i=2 : \bar C_{20} &=& \{0000000, 0011010, 0001101, 0010111\} \nonumber \\
\bar C_{21} &=& \{1010001, 1001011, 1011100, 1000110\} \nonumber \\
\bar C_{22} &=& \{1101000, 1110010, 1100101, 1111111\} \nonumber \\
\bar C_{23} &=& \{0111001, 0100011, 0110100, 0101110\} \nonumber \\
i=3 : \bar C_{30} &=& \{0000000, 1010001, 0001101, 1011100\} \nonumber \\
\bar C_{31} &=& \{0011010, 1001011, 0010111, 1000110\} \nonumber \\
\bar C_{32} &=& \{1101000, 0111001, 1100101, 0110100\} \nonumber \\
\bar C_{33} &=& \{1110010, 0100011, 1111111, 0101110\} \nonumber \\
i=4 : \bar C_{40} &=& \{0000000, 1010001, 1101000, 0111001\} \nonumber \\
\bar C_{41} &=& \{0011010, 1001011, 1110010, 0100011\} \nonumber \\
\bar C_{42} &=& \{0001101, 1011100, 1100101, 0110100\} \nonumber \\
\bar C_{43} &=& \{0010111, 1000110, 1111111, 0101110\} \nonumber \\
i=5 : \bar C_{50} &=& \{0000000, 1010001, 1101000, 0111001\} \nonumber \\
\bar C_{51} &=& \{0011010, 1001011, 1110010, 0100011\} \nonumber \\
\bar C_{52} &=& \{0001101, 1011100, 1100101, 0110100\} \nonumber \\
\bar C_{53} &=& \{0010111, 1000110, 1111111, 0101110\} \nonumber \\
i=6 : \bar C_{60} &=& \{0000000, 1010001, 0011010, 1001011, \nonumber \\
&& 1101000, 0111001, 1110010, 0100011\} \nonumber \\
\bar C_{61} &=& \{0001101, 1011100, 0010111, 1000110, \nonumber \\
&& 1100101, 0110100, 1111111, 0101110\} . \nonumber
\end{eqnarray}
\par
The resulting tail-biting trellis is shown in Fig.5. A modified tail-biting trellis obtained using state permutations is shown in Fig.6. We observe that the trellis in Fig.6 is identical to the one in Fig.4.
\begin{figure}[tb]
\begin{center}
\includegraphics[width=8.0cm,clip]{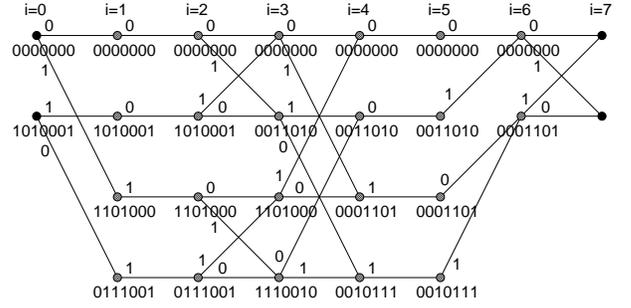}
\end{center}
\caption{The tail-biting trellis obtained using the proposed construction.}
\label{fig:5}
\end{figure}
\begin{figure}[tb]
\begin{center}
\includegraphics[width=8.0cm,clip]{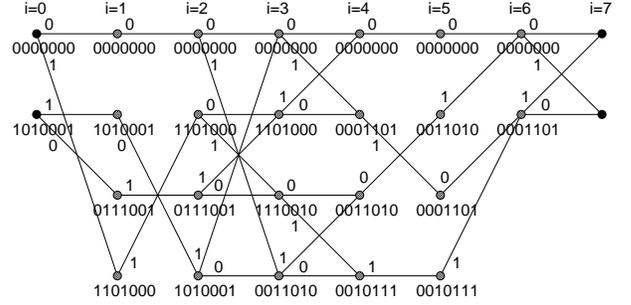}
\end{center}
\caption{A modified tail-biting trellis obtained using state permutations.}
\label{fig:6}
\end{figure}

\subsection{Algebraic construction of a BCJR-dual trellis}
In this section, we consider the BCJR-dual trellis $T^{\perp}=T_{(H, G, \Theta^T)}$~\cite{glu 112, nori 06} corresponding to a tail-biting BCJR trellis $T_{(G, H, \Theta)}$. Note that the state matrix $\hat N_i$ associated with $T^{\perp}$ is given by $\hat N_i=N_i^T$~\cite{nori 06}. Let $\hat V_i$ be the state space of $T^{\perp}$ at level $i$. We have
\begin{equation}
\hat V_i=\mbox{im}\hat N_i\cong \mbox{F}^{n-k}/\mbox{ker}(\hat N_i),~\mbox{for}~i=0, \cdots, n-1.
\end{equation}
Based on this equation, we can construct a tail-biting trellis which is isomorphic to the tail-biting BCJR-dual trellis.
\par
{\it Example 1 (Continued):} Again, consider Example 1. We only need to compute $\mbox{ker}(\hat N_i)$ after obtaining the state matrices $\hat N_i=N_i^T$. For $i=0, \cdots, n-1$, let $\bar C_{i0}=\mbox{ker}(\hat N_i)$. Then a tail-biting trellis is constructed based on the partitions $C/\bar C_{i0}$. The resulting tail-biting trellis is shown in Fig.7. It is shown that the obtained trellis is isomorphic to the tail-biting BCJR-dual trellis $T_{(H, G, \Theta^T)}$.
\begin{figure}[tb]
\begin{center}
\includegraphics[width=8.0cm,clip]{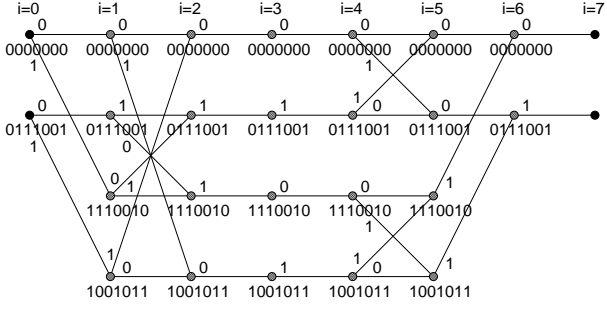}
\end{center}
\caption{The tail-biting BCJR-dual trellis obtained using the proposed construction.}
\label{fig:7}
\end{figure}

\subsection{Algebraic construction of a KV trellis}
Consider an $(n, k)$ linear block code $C$. Let $X$ be a characteristic matrix~\cite{koe 03} associated with $C$. We set
\begin{equation}
G=\left(
\begin{array}{c}
\mbox{\boldmath $g$}_1 \\
\mbox{\boldmath $g$}_2 \\
\cdots \\
\mbox{\boldmath $g$}_k
\end{array}
\right)
\begin{array}{c}
(a_1, b_1] \\
(a_2, b_2] \\
\cdots \\
(a_k, b_k] ,
\end{array}
\end{equation}
where $\mbox{\boldmath $g$}_1, \cdots \mbox{\boldmath $g$}_k$ are $k$ linearly independent rows of $X$. Consider the KV trellis, denoted by $T_{G, S}$, constructed based on $G$ and its span list $S=[(a_1, b_1], \cdots, (a_k, b_k]]$. According to~\cite{glu 111,glu 112}, the state matrix at level $i$ is given by
\begin{eqnarray}
M_i &=& \left(
\begin{array}{ccc}
\mu_i^1 & &  \\
 & \ddots & \\
 & & \mu_i^k
\end{array}
\right) \in \mbox{F}^{k \times k} \\
\mu_i^l &=& \left\{
\begin{array}{ll}
1, & \mbox{if}~i \in (a_l, b_l], \\
0, & \mbox{if}~i \notin (a_l, b_l] .
\end{array} \right.
\end{eqnarray}
Also, the state space at level $i$ is expressed as
\begin{equation}
V_i=\mbox{im}M_i,~\mbox{for}~i=0, \cdots, n-1.
\end{equation}
Using the homomorphism theorem, the right-hand side becomes
\begin{equation}
\mbox{im}M_i\cong \mbox{F}^k/\mbox{ker}(M_i),~\mbox{for}~i=0, \cdots, n-1,
\end{equation}
where $\mbox{ker}(M_i)$ is the kernel of the linear mapping induced by $M_i$. Then we have
\begin{equation}
V_i=\mbox{im}M_i\cong \mbox{F}^k/\mbox{ker}(M_i),~\mbox{for}~i=0, \cdots, n-1.
\end{equation}
For $i=0, \cdots, n-1$, let
\begin{equation}
\mbox{F}^k/\mbox{ker}(M_i)=\{\bar C_{i0}, \bar C_{i1}, \cdots, \bar C_{i,m(i)-1}\} .
\end{equation}
Here it is assumed that the elements in $\bar C_{il}$ have been transformed into the corresponding codewords. In this case, a tail-biting trellis is constructed by tracing the cosets $\bar C_{il}$ which contain {\boldmath $c$} for each codeword $\mbox{\boldmath $c$}\in C$.
\par
For the obtaind tail-biting trellis, we have the following.
\begin{pro}
A tail-biting trellis obtained using the proposed construction, denoted by $T_{alg}$, is isomorphic to the associated KV trellis $T_{G, S}$. If the coset representatives (in terms of $\mbox{\boldmath $u$}$) are placed in ascending order at each level $i$, then $T_{alg}$ is identical to $T_{G, S}$.
\end{pro}
\begin{IEEEproof}
The proposed method is based on the isomorphism:
\begin{displaymath}
\mbox{im}M_i\cong \mbox{F}^k/\mbox{ker}(M_i),~\mbox{for}~i=0, \cdots, n-1.
\end{displaymath}
For $\mbox{\boldmath $u$},~\mbox{\boldmath $u$}' \in \mbox{F}^k$, let $\mbox{\boldmath $c$}$ and $\mbox{\boldmath $c$}'$ be the corresponding codewords. Suppose that $\mbox{\boldmath $u$}-\mbox{\boldmath $u$}' \in \mbox{ker}(M_i)$. This means that $\mbox{\boldmath $c$}$ and $\mbox{\boldmath $c$}'$ are contained in the same coset, i.e., go through the same state in the trellis at level $i$. On the other hand, $\mbox{\boldmath $u$}-\mbox{\boldmath $u$}' \in \mbox{ker}(M_i)$ is equivalent to $\mbox{\boldmath $u$}M_i=\mbox{\boldmath $u$}'M_i$. We have two cases:
\par
1) $\mu_i^l=1$: In this case, since
\begin{eqnarray}
\mbox{\boldmath $u$}M_i &=& (\cdots, u_l, \cdots) \nonumber \\
\mbox{\boldmath $u$}'M_i &=& (\cdots, u_l', \cdots) , \nonumber
\end{eqnarray}
it follows that $u_l=u_l'$.
\par
2) $\mu_i^l=0$: In this case, we have
\begin{eqnarray}
\mbox{\boldmath $u$}M_i &=& (\cdots, 0, \cdots) \nonumber \\
\mbox{\boldmath $u$}'M_i &=& (\cdots, 0, \cdots) \nonumber
\end{eqnarray}
regardless of the values of $u_l$ and $u_l'$. Hence, only $u_l$ such that $\mu_i^l=1$ are effective. That is, $\mbox{\boldmath $u$}$ and $\mbox{\boldmath $u$}'$ whose components coincide at positions $l$ such that $\mu_i^l=1$ are contained in the same coset.
\par
On the other hand, a KV trellis is obtained as the product of elementary trellises $T_{g_l}$. Let $i$ be any level in the trellis. If $\mu_i^l=1$ holds, then $T_{g_l}$ has ``two'' vertices at level $i$ from the definition of an elementary trellis~\cite{koe 03}. Hence, both $0$ and $1$ are allowed as the value of $u_l$ in the product of elementary trellises. This is equivalent to considering only the components $u_l$ such that $\mu_i^l=1$.
\par
Next, consider the transition from level $(i-1)$ to level $i$. Without loss of generality, suppose that $\mu_{i-1}^l=0,~\mu_i^l=1$.
\par
First, consider the ``algebraic'' construction. Since $\mu_i^l=1$, $u_l=0$ and $u_l=1$ are distinguished at level $i$. Hence, denote by $\bar U_i(u_l=0)$ and $\bar U_i(u_l=1)$ the cosets which contain $\mbox{\boldmath $u$}=(\cdots, u_l=0, \cdots)$ and $\mbox{\boldmath $u$}'=(\cdots, u_l=1, \cdots)$, respectively. On the other hand, since $\mu_{i-1}^l=0$, $u_l$ is not effective at level $i-1$. That is, $\mbox{\boldmath $u$}=(\cdots, u_l=0, \cdots)$ and $\mbox{\boldmath $u$}'=(\cdots, u_l=1, \cdots)$ are contained in the same coset $\bar U_{i-1}(u_l=0)$. Noting these facts, if $u_l=0$ at level $i$, then we regard the value of $u_l$ at level $i-1$ as $0$, whereas if $u_l=1$ at level $i$, then we regard the value of $u_l$ at level $i-1$ as $1$. This means that there exist transitions: $\bar U_{i-1}(u_l=0)\rightarrow \bar U_i(u_l=0)$ and $\bar U_{i-1}(u_l=0)\rightarrow \bar U_i(u_l=1)$.
\par
Next, consider the ``product'' construction. From the definition of an elementary trellis, our assumption ($\mu_{i-1}^l=0,~\mu_i^l=1$) corresponds to the left end of the span $(a_l, b_l]$ (i.e., $i-1=a_l$ and $i=a_l+1$). Then there are two transitions: $u_l=0~(i-1)\rightarrow u_l=0~(i)$ and $u_l=0~(i-1)\rightarrow u_l=1~(i)$. This is equivalent to the above. Moreover, the edge label from $(i-1)$ to $i$ in $T_{g_l}$ is defined as $\beta \cdot g_{l,i-1}$~\cite[Section IV-C]{koe 03}. Since $\mu_i^l=1$, $\beta$ is equal to $u_l$ from the definition of $\beta$. Here note that $u_l\cdot g_{l, i-1}$ is the $l$th component of the product $\mbox{\boldmath $u$}\cdot \mbox{\boldmath $\bar g$}_{i-1}$, where $\mbox{\boldmath $\bar g$}_{i-1}$ is the $(i-1)$th column of $G$.
\end{IEEEproof}
\par
{\it Example 2 (Nori and Shankar~\cite{nori 06}):} Consider the linear block code $C$ defined by
\begin{equation}
G=\left(
\begin{array}{ccccccc}
0 & 0 & 0 & 1 & 1 & 0 & 1 \\
1 & 1 & 0 & 1 & 0 & 0 & 0 \\
0 & 0 & 1 & 1 & 0 & 1 & 0 \\
1 & 0 & 1 & 0 & 0 & 0 & 1 
\end{array}
\right) .
\end{equation}
A characteristic matrix associated with $C$ is given by
\begin{equation}
X=\left(
\begin{array}{ccccccc}
\mbox{\boldmath $1$}& \mbox{\boldmath $1$} & \mbox{\boldmath $0$} & \mbox{\boldmath $1$} & 0 & 0 & 0 \\
0 & \mbox{\boldmath $1$}& \mbox{\boldmath $1$} & \mbox{\boldmath $0$} & \mbox{\boldmath $1$} & 0 & 0 \\
0& 0 & \mbox{\boldmath $1$} & \mbox{\boldmath $1$} & \mbox{\boldmath $0$} & \mbox{\boldmath $1$} & 0 \\
0& 0 & 0 & \mbox{\boldmath $1$} & \mbox{\boldmath $1$} & \mbox{\boldmath $0$} & \mbox{\boldmath $1$} \\
\mbox{\boldmath $1$} & 0 & 0 & 0 & \mbox{\boldmath $1$} & \mbox{\boldmath $1$} & \mbox{\boldmath $0$} \\
\mbox{\boldmath $0$} & \mbox{\boldmath $1$} & 0 & 0 & 0 & \mbox{\boldmath $1$} & \mbox{\boldmath $1$} \\
\mbox{\boldmath $1$}& \mbox{\boldmath $0$} & \mbox{\boldmath $1$} & 0 & 0 & 0 & \mbox{\boldmath $1$}
\end{array}
\right)
\begin{array}{l}
$(0, 3]$ \\
$(1, 4]$ \\
$(2, 5]$ \\
$(3, 6]$ \\
$(4, 0]$ \\
$(5, 1]$ \\
$(6, 2]$ .
\end{array}
\end{equation}
\par
By selecting $4$ linearly independent rows of $X$, we set
\begin{equation}
G=\left(
\begin{array}{ccccccc}
0& 0 & 0 & \mbox{\boldmath $1$} & \mbox{\boldmath $1$} & \mbox{\boldmath $0$} & \mbox{\boldmath $1$} \\
\mbox{\boldmath $1$}& \mbox{\boldmath $1$} & \mbox{\boldmath $0$} & \mbox{\boldmath $1$} & 0 & 0 & 0 \\
0& 0 & \mbox{\boldmath $1$} & \mbox{\boldmath $1$} & \mbox{\boldmath $0$} & \mbox{\boldmath $1$} & 0 \\
\mbox{\boldmath $1$}& \mbox{\boldmath $0$} & \mbox{\boldmath $1$} & 0 & 0 & 0 & \mbox{\boldmath $1$}
\end{array}
\right)
\begin{array}{l}
$(3, 6]$ \\
$(0, 3]$ \\
$(2, 5]$ \\
$(6, 2]$ .
\end{array}
\end{equation}
The KV trellis $T_{G, S}$ based on $G$ and its span list $S$ is shown in Fig.8.
\begin{figure}[tb]
\begin{center}
\includegraphics[width=8.0cm,clip]{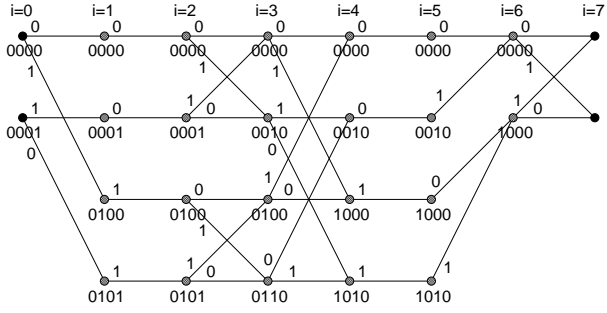}
\end{center}
\caption{The KV trellis obtained using the product construction.}
\label{fig:8}
\end{figure}
\par
In order to apply the proposed construction, we first compute the state matrices $M_i~(0\leq i \leq 6)$. They are given as follows:
\begin{eqnarray}
M_0 &=& \left(
\begin{array}{cccc}
0& 0 & 0 & 0 \\
0& 0 & 0 & 0 \\
0& 0 & 0 & 0 \\
0& 0 & 0 & 1 
\end{array}
\right) \nonumber \\
M_1 &=& \left(
\begin{array}{cccc}
0& 0 & 0 & 0 \\
0& 1 & 0 & 0 \\
0& 0 & 0 & 0 \\
0& 0 & 0 & 1
\end{array}
\right) \nonumber \\
M_2 &=& \left(
\begin{array}{cccc}
0& 0 & 0 & 0 \\
0& 1 & 0 & 0 \\
0& 0 & 0 & 0 \\
0& 0 & 0 & 1
\end{array}
\right) \nonumber \\
M_3 &=& \left(
\begin{array}{cccc}
0& 0 & 0 & 0 \\
0& 1 & 0 & 0 \\
0& 0 & 1 & 0 \\
0& 0 & 0 & 0
\end{array}
\right) \nonumber \\
M_4 &=& \left(
\begin{array}{cccc}
1& 0 & 0 & 0 \\
0& 0 & 0 & 0 \\
0& 0 & 1 & 0 \\
0& 0 & 0 & 0
\end{array}
\right) \nonumber \\
M_5 &=& \left(
\begin{array}{cccc}
1& 0 & 0 & 0 \\
0& 0 & 0 & 0 \\
0& 0 & 1 & 0 \\
0& 0 & 0 & 0
\end{array}
\right) \nonumber \\
M_6 &=& \left(
\begin{array}{cccc}
1& 0 & 0 & 0 \\
0& 0 & 0 & 0 \\
0& 0 & 0 & 0 \\
0& 0 & 0 & 0
\end{array}
\right) . \nonumber
\end{eqnarray}
Next, $\mbox{ker}(M_i)~(0\leq i \leq 6)$ are obtained as follows:
\begin{eqnarray}
\mbox{ker}(M_0) &=& \{0000, 0010, 0100, 0110, \nonumber \\
&& 1000, 1010, 1100, 1110\} \nonumber \\
\mbox{ker}(M_1) &=& \{0000, 0010, 1000, 1010\} \nonumber \\
\mbox{ker}(M_2) &=& \{0000, 0010, 1000, 1010\} \nonumber \\
\mbox{ker}(M_3) &=& \{0000, 0001, 1000, 1001\} \nonumber \\
\mbox{ker}(M_4) &=& \{0000, 0001, 0100, 0101\} \nonumber \\
\mbox{ker}(M_5) &=& \{0000, 0001, 0100, 0101\} \nonumber \\
\mbox{ker}(M_6) &=& \{0000, 0001, 0010, 0011, \nonumber \\
&& 0100, 0101, 0110, 0111\} . \nonumber 
\end{eqnarray}
\par
{\it Remark 4:} Note that $\mbox{ker}(M_i)=\mbox{ker}(N_i)~(0\leq i \leq 6)$ holds, where $\mbox{ker}(N_i)$ are the kernels in Example 1. We can show this equality under a general condition.
\par
Then we have the partitions $\mbox{F}^4/\mbox{ker}(M_i)$:
\begin{eqnarray}
i=0 : \bar U_{00} &=& \{0000, 0010, 0100, 0110, \nonumber \\
&& 1000, 1010, 1100, 1110\} \nonumber \\
\bar U_{01} &=& \{0001, 0011, 0101, 0111, \nonumber \\
&& 1001, 1011, 1101, 1111\} \nonumber \\
i=1 : \bar U_{10} &=& \{0000, 0010, 1000, 1010\} \nonumber \\
\bar U_{11} &=& \{0001, 0011, 1001, 1011\} \nonumber \\
\bar U_{12} &=& \{0100, 0110, 1100, 1110\} \nonumber \\
\bar U_{13} &=& \{0101, 0111, 1101, 1111\} \nonumber \\
i=2 : \bar U_{20} &=& \{0000, 0010, 1000, 1010\} \nonumber \\
\bar U_{21} &=& \{0001, 0011, 1001, 1011\} \nonumber \\
\bar U_{22} &=& \{0100, 0110, 1100, 1110\} \nonumber \\
\bar U_{23} &=& \{0101, 0111, 1101, 1111\} \nonumber \\
i=3 : \bar U_{30} &=& \{0000, 0001, 1000, 1001\} \nonumber \\
\bar U_{31} &=& \{0010, 0011, 1010, 1011\} \nonumber \\
\bar U_{32} &=& \{0100, 0101, 1100, 1101\} \nonumber \\
\bar U_{33} &=& \{0110, 0111, 1110, 1111\} \nonumber \\
i=4 : \bar U_{40} &=& \{0000, 0001, 0100, 0101\} \nonumber \\
\bar U_{41} &=& \{0010, 0011, 0110, 0111\} \nonumber \\
\bar U_{42} &=& \{1000, 1001, 1100, 1101\} \nonumber \\
\bar U_{43} &=& \{1010, 1011, 1110, 1111\} \nonumber \\
i=5 : \bar U_{50} &=& \{0000, 0001, 0100, 0101\} \nonumber \\
\bar U_{51} &=& \{0010, 0011, 0110, 0111\} \nonumber \\
\bar U_{52} &=& \{1000, 1001, 1100, 1101\} \nonumber \\
\bar U_{53} &=& \{1010, 1011, 1110, 1111\} \nonumber \\
i=6 : \bar U_{60} &=& \{0000, 0001, 0010, 0011, \nonumber \\
&& 0100, 0101, 0110, 0111\} \nonumber \\
\bar U_{61} &=& \{1000, 1001, 1010, 1011, \nonumber \\
&& 1100, 1101, 1110, 1111\} . \nonumber
\end{eqnarray}
\par
{\it Remark 5:} We observe that at each level $i$, the representatives of the cosets coincide with the states in the trellis in Fig.8.
\par
Since we have seen that $\mbox{ker}(M_i)=\mbox{ker}(N_i)~(0\leq i \leq 6)$, the same tail-biting trellis as that in Fig.8. is obtained. The resulting trellis is shown in Fig.9.
\par
As is stated above, we have the following.
\newtheorem{lem}{Lemma}
\begin{lem}
\begin{equation}
\mbox{ker}(M_i)=\mbox{ker}(N_i),~\mbox{for}~i=0, \cdots, n-1 .
\end{equation}
\end{lem}
\begin{IEEEproof}
Suppose that $\mu_i^l=0$, i.e., $i \notin (a_l, b_l]$. That is, the $l$th row of $M_i$ is the all-zero vector. We first show that the $l$th row of $N_i$, denoted by $N_{il}$, is also the all-zero vector. Without loss of generality, let $(a_l, b_l]~(a_l>b_l)$ be a circular span. From the assumption that $i \notin (a_l, b_l]$, it follows that $b_l+1 \leq i \leq a_l$. In this case, $N_{il}$ is expressed as
\begin{displaymath}
N_{il}=g_{l,0}\mbox{\boldmath $h$}_0^T+g_{l,1}\mbox{\boldmath $h$}_1^T+\cdots +g_{l,b_l}\mbox{\boldmath $h$}_{b_l}^T+\mbox{\boldmath $d$}_{g_l} .
\end{displaymath}
Here note that
\begin{displaymath}
\mbox{\boldmath $d$}_{g_l}=g_{l,a_l}\mbox{\boldmath $h$}_{a_l}^T+\cdots +g_{l,n-1}\mbox{\boldmath $h$}_{n-1}^T .
\end{displaymath}
Then we have
\begin{eqnarray}
N_{il} &=& g_{l,0}\mbox{\boldmath $h$}_0^T++\cdots +g_{l,b_l}\mbox{\boldmath $h$}_{b_l}^T \nonumber \\
&& +g_{l,a_l}\mbox{\boldmath $h$}_{a_l}^T+\cdots +g_{l,n-1}\mbox{\boldmath $h$}_{n-1}^T . \nonumber
\end{eqnarray}
Since $\mbox{\boldmath $g$}_l$ is a codeword, $N_{il}$ is the all-zero vector.
\par
On the other hand, it has been shown that $\mbox{rank}M_i=\mbox{rank}N_i$ (see~\cite{glu 111}). This implies that any two non-zero rows of $N_i$ are different. Then it follows that $\mbox{ker}(M_i)=\mbox{ker}(N_i)$.
\end{IEEEproof}
\par
Suppose that a KV trellis $T_{G, S}$ and the associated tail-biting BCJR trellis $T_{(G, H, \Theta)}$ are constructed using the proposed algebraic method. Since $\mbox{ker}(M_i)=\mbox{ker}(N_i)$ holds, the two resulting trellises are identical. Hence, we have shown the following.
\begin{pro}
$T_{G, S}$ and $T_{(G, H, \Theta)}$ are isomorphic.
\end{pro}
\par
Note that this is the result in~\cite[Theorem IV.11]{glu 111}.
\begin{figure}[tb]
\begin{center}
\includegraphics[width=8.0cm,clip]{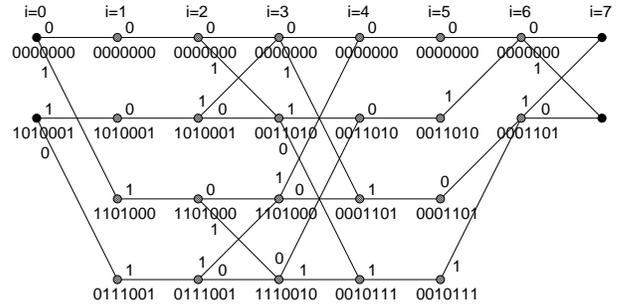}
\end{center}
\caption{The tail-biting trellis obtained using the proposed construction.}
\label{fig:9}
\end{figure}

% end of file

%% file: sec.4-d.tex
\section{Complexity of Tail-Biting Trellises Obtained Using the Proposed Construction}
In this section, we discuss the complexity of the tail-biting trellises obtained using the proposed construction. We have the following.
\begin{pro}
\begin{equation}
\vert V_i \vert=2^{k-\mbox{dim ker}(N_i)},~\mbox{for}~i=0, \cdots, n-1.
\end{equation}
\end{pro}
\begin{IEEEproof}
From $V_i \cong \mbox{F}^k/\mbox{ker}(N_i)$, we have
\begin{displaymath}
\mbox{dim} V_i=k-\mbox{dim ker}(N_i) .
\end{displaymath}
\end{IEEEproof}
\par
A similar result holds for the edge spaces $E_i$.
\begin{pro}
\begin{equation}
\vert E_i \vert=2^{k-\mbox{dim ker}(N_{i-1}, \mbox{\boldmath $\bar g$}_i, N_i)},~\mbox{for}~i=1, \cdots, n.
\end{equation}
\end{pro}
\begin{IEEEproof}
From $E_i \cong \mbox{F}^k/\mbox{ker}(N_{i-1}, \mbox{\boldmath $\bar g$}_i, N_i)$, we have
\begin{displaymath}
\mbox{dim} E_i=k-\mbox{dim ker}(N_{i-1}, \mbox{\boldmath $\bar g$}_i, N_i).
\end{displaymath}
\end{IEEEproof}
\newtheorem{cor}{Corollary}
\begin{cor} Denote by $\rho_i^+$ and $\rho_i^-$ the out-degree and in-degree at level $i$, respectively. Then we have
\begin{equation}
\rho_i^+=2^{\mbox{dim ker}(N_i)-\mbox{dim ker}(N_i, \mbox{\boldmath $\bar g$}_{i+1}, N_{i+1})}
\end{equation}
\begin{equation}
\rho_i^-=2^{\mbox{dim ker}(N_i)-\mbox{dim ker}(N_{i-1}, \mbox{\boldmath $\bar g$}_i, N_i)} .
\end{equation}
\end{cor}
\begin{IEEEproof}
The edge space at level $i+1$ is defined as
\begin{displaymath}
E_{i+1}\cong \mbox{F}^k/\mbox{ker}(N_i, \mbox{\boldmath $\bar g$}_{i+1}, N_{i+1}).
\end{displaymath}
Let $v \in V_i$ be any state (i.e., $v \in \mbox{F}^k/\mbox{ker}(N_i)$). Let us denote the set of edges $\in E_{i+1}$ leaving $v$ by $E_{i+1}^v$. Then  $E_{i+1}^v$ is the set of cosets $e \in \mbox{F}^k/\mbox{ker}(N_i, \mbox{\boldmath $\bar g$}_{i+1}, N_{i+1})$ such that $e \subset v$. Hence, each of the sets $E_{i+1}^v$ has the same size. Then the former result follows from the relation:
\begin{displaymath}
\rho_i^+=\frac{\vert E_{i+1} \vert}{\vert V_i \vert}.
\end{displaymath}
The proof of the latter equality is similar.
\end{IEEEproof}
\par
Next, consider the BCJR-dual trellis $T^{\perp}$ corresponding to a tail-biting BCJR trellis $T$. Let $\hat V_i$ and $\hat E_i$ be the associated state and edge spaces of $T^{\perp}$, respectively. As in the case of a tail-biting BCJR trellis,
\begin{eqnarray}
\hat V_i &=& \mbox{im}\hat N_i \nonumber \\
&\cong& \mbox{F}^{n-k}/\mbox{ker}(\hat N_i),~\mbox{for}~i=0, \cdots, n-1 \nonumber \\
\hat E_i &=& \mbox{im}(\hat N_{i-1}, \mbox{\boldmath $h$}_i, \hat N_i) \nonumber \\
&\cong& \mbox{F}^{n-k}/\mbox{ker}(\hat N_{i-1}, \mbox{\boldmath $h$}_i, \hat N_i),~\mbox{for}~i=1, \cdots, n \nonumber
\end{eqnarray}
hold. We have the following.
\begin{pro}
\begin{equation}
\vert V_i \vert =\vert \hat V_i \vert,~\mbox{for}~i=0, \cdots, n-1.
\end{equation}
\end{pro}
\begin{IEEEproof}
Note that
\begin{eqnarray}
\mbox{dim} V_i &=& \mbox{dim im}N_i=\mbox{rank}N_i \nonumber \\
\mbox{dim} \hat V_i &=& \mbox{dim im}\hat N_i=\mbox{rank}\hat N_i . \nonumber
\end{eqnarray}
Here $\mbox{rank}N_i=\mbox{rank}N_i^T=\mbox{rank}\hat N_i$ holds.
\end{IEEEproof}
\par
Next, consider the relation between $\mbox{dim ker}(N_{i-1}, \mbox{\boldmath $\bar g$}_i, N_i)$ and $\mbox{dim ker}(\hat N_{i-1}, \mbox{\boldmath $h$}_i, \hat N_i)$. For the purpose, we impose some restrictions.
\par
In~\cite{glu 112}, Gluesing-Luerssen and Weaver showed that for each complete set of characteristic generators of a code, there exists a complete set of characteristic generators of the dual code such that their resulting KV trellises are dual to each other if paired suitably. Hence, if $T$ is a KV trellis, then there exists a KV trellis $\hat T$ which is dual to $T$.
\par
{\it Example 3:} Again consider the $(7, 4)$ Hamming code $C$ defined by
\begin{displaymath}
G=\left(
\begin{array}{ccccccc}
0& 0 & 0 & \mbox{\boldmath $1$} & \mbox{\boldmath $1$} & \mbox{\boldmath $0$} & \mbox{\boldmath $1$} \\
\mbox{\boldmath $1$}& \mbox{\boldmath $1$} & \mbox{\boldmath $0$} & \mbox{\boldmath $1$} & 0 & 0 & 0 \\
0& 0 & \mbox{\boldmath $1$} & \mbox{\boldmath $1$} & \mbox{\boldmath $0$} & \mbox{\boldmath $1$} & 0 \\
\mbox{\boldmath $1$}& \mbox{\boldmath $0$} & \mbox{\boldmath $1$} & 0 & 0 & 0 & \mbox{\boldmath $1$}
\end{array}
\right)
\begin{array}{l}
$(3, 6]$ \\
$(0, 3]$ \\
$(2, 5]$ \\
$(6, 2]$ .
\end{array}
\end{displaymath}
A characteristic matrix $Y$ associated with $C^{\perp}$ is given by
\begin{equation}
Y=\left(
\begin{array}{ccccccc}
\mbox{\boldmath $1$}& \mbox{\boldmath $0$} & \mbox{\boldmath $1$} & \mbox{\boldmath $1$} & \mbox{\boldmath $1$} & 0 & 0 \\
0 & \mbox{\boldmath $1$}& \mbox{\boldmath $0$} & \mbox{\boldmath $1$} & \mbox{\boldmath $1$} & \mbox{\boldmath $1$} & 0 \\
0& 0 & \mbox{\boldmath $1$} & \mbox{\boldmath $0$} & \mbox{\boldmath $1$} & \mbox{\boldmath $1$} & \mbox{\boldmath $1$} \\
\mbox{\boldmath $1$} & 0 & 0 & \mbox{\boldmath $1$} & \mbox{\boldmath $0$} & \mbox{\boldmath $1$} & \mbox{\boldmath $1$} \\
\mbox{\boldmath $1$} & \mbox{\boldmath $1$} & 0  & 0 & \mbox{\boldmath $1$} & \mbox{\boldmath $0$} & \mbox{\boldmath $1$} \\
\mbox{\boldmath $1$} & \mbox{\boldmath $1$} & \mbox{\boldmath $1$} & 0 & 0 & \mbox{\boldmath $1$} & \mbox{\boldmath $0$} \\
\mbox{\boldmath $0$}& \mbox{\boldmath $1$} & \mbox{\boldmath $1$} & \mbox{\boldmath $1$} & 0 & 0 & \mbox{\boldmath $1$}
\end{array}
\right)
\begin{array}{l}
$(0, 4]$ \\
$(1, 5]$ \\
$(2, 6]$ \\
$(3, 0]$ \\
$(4, 1]$ \\
$(5, 2]$ \\
$(6, 3]$ .
\end{array}
\end{equation}
By selecting $3$ linearly independent rows of $Y$, we set
\begin{equation}
\hat H=\left(
\begin{array}{ccccccc}
\mbox{\boldmath $1$}& \mbox{\boldmath $0$} & \mbox{\boldmath $1$} & \mbox{\boldmath $1$} & \mbox{\boldmath $1$} & 0 & 0 \\
0 & \mbox{\boldmath $1$}& \mbox{\boldmath $0$} & \mbox{\boldmath $1$} & \mbox{\boldmath $1$} & \mbox{\boldmath $1$} & 0 \\
\mbox{\boldmath $1$} & \mbox{\boldmath $1$} & 0  & 0 & \mbox{\boldmath $1$} & \mbox{\boldmath $0$} & \mbox{\boldmath $1$} 
\end{array}
\right)
\begin{array}{l}
$(0, 4]$ \\
$(1, 5]$ \\
$(4, 1]$ .
\end{array}
\end{equation}
Note that $(G, \hat H)$ is a ``dual selection'' of $(X, Y)$~\cite{glu 112}. We see that the KV trellises $T_{G, S}$ and $T_{\hat H, \hat S}$ are dual to each other. This can also be stated in terms of BCJR representations. That is, $T_{(\hat H, G, \hat S)}=T_{(G, \hat H, S)}^{\perp}$ holds~\cite{glu 112}.
\par
Taking account of the above argument, we first show the following lemma.
\begin{lem}
Suppose that a KV trellis $T$ and its dual $\hat T$ are constructed based on extended minimal-span generator matrices (e-MSGM's) (see Section V). Also, let $(\alpha_i, \beta_i)$ and $(\hat \alpha_i, \hat \beta_i)$ be the variables for $T$ and $\hat T$, respectively. In this case, we have
\begin{eqnarray}
\alpha_i &=& -\hat \alpha_i+\hat \beta_{i-1}+\hat \beta_i+1,~\mbox{for}~i=1, \cdots, n \\
\hat \alpha_i &=& -\alpha_i+\beta_{i-1}+\beta_i+1,~\mbox{for}~i=1, \cdots, n .
\end{eqnarray}
\end{lem}
\begin{IEEEproof}
See Appendix.
\end{IEEEproof}
\par
Now suppose that KV trellises $T_{G, S}$ and $T_{\hat H, \hat S}$ are dual to each other. Let $H$ be a parity-check matrix corresponding to $G$. Then we have the following.
\begin{eqnarray}
T_{G, S} &\cong& T_{(G, H, S)} \\
T_{\hat H, \hat S} &\cong& T_{(G, H, S)}^{\perp}.
\end{eqnarray}
Here the latter equation is derived from the relations $T_{(\hat H, G, \hat S)}=T_{(G, \hat H, S)}^{\perp}$ and $T_{(G, \hat H, S)}^{\perp}\cong T_{(G, H, S)}^{\perp}$. Let $N_i$ and $\hat N_i~(=N_i^T)$ be the state matrices of $T_{(G, H, S)}$ and $T_{(G, H, S)}^{\perp}$, respectively. Then we have the following.
\begin{pro}
\begin{eqnarray}
\lefteqn{k-\mbox{dim ker}(N_{i-1}, \mbox{\boldmath $\bar g$}_i, N_i)} \nonumber \\
&& =(n-k+1)+\mbox{dim ker}(\hat N_{i-1}, \mbox{\boldmath $h$}_i, \hat N_i) \nonumber \\
&& \quad -\mbox{dim ker}(\hat N_{i-1})-\mbox{dim ker}(\hat N_i)
\end{eqnarray}
\begin{eqnarray}
\lefteqn{(n-k)-\mbox{dim ker}(\hat N_{i-1}, \mbox{\boldmath $h$}_i, \hat N_i)} \nonumber \\
&& =(k+1)+\mbox{dim ker}(N_{i-1}, \mbox{\boldmath $\bar g$}_i, N_i) \nonumber \\
&& \quad -\mbox{dim ker}(N_{i-1})-\mbox{dim ker}(N_i) .
\end{eqnarray}
\end{pro}
\begin{IEEEproof}
The number of vertices at level $i$ is $\vert E_i \vert =2^{\alpha_i}$ (see Appendix). Using Lemma 2, we have
\begin{displaymath}
\vert E_i \vert=2^{-\hat \alpha_i+\hat \beta_{i-1}+\hat \beta_i+1} .
\end{displaymath}
Also (cf. Appendix), it follows that
\begin{eqnarray}
\hat \alpha_i &=& (n-k)-\mbox{dim ker}(\hat N_{i-1}, \mbox{\boldmath $h$}_i, \hat N_i) \nonumber \\
\hat \beta_{i-1} &=& (n-k)-\mbox{dim ker}(\hat N_{i-1}) \nonumber \\
\hat \beta_i &=& (n-k)-\mbox{dim ker}(\hat N_i) . \nonumber
\end{eqnarray}
Then we have
\begin{eqnarray*}
\lefteqn{-\hat \alpha_i+\hat \beta_{i-1}+\hat \beta_i+1} \\
&& =-(n-k)+\mbox{dim ker}(\hat N_{i-1}, \mbox{\boldmath $h$}_i, \hat N_i) \\
&& \quad +(n-k)-\mbox{dim ker}(\hat N_{i-1}) \\
&& \quad +(n-k)-\mbox{dim ker}(\hat N_i)+1 \\
&& =(n-k+1)+\mbox{dim ker}(\hat N_{i-1}, \mbox{\boldmath $h$}_i, \hat N_i) \\
&& \quad -\mbox{dim ker}(\hat N_{i-1})-\mbox{dim ker}(\hat N_i) .
\end{eqnarray*}
On the other hand, we know that
\begin{displaymath}
\vert E_i \vert =2^{k-\mbox{dim ker}(N_{i-1}, \mbox{\boldmath $\bar g$}_i, N_i)} .
\end{displaymath}
Then the result follows. The second equation is derived in a similar way.
\end{IEEEproof}
\par
{\it Example 4:} Consider the tail-biting BCJR trellis in Example 1 and its BCJR-dual trellis in Example 1 (Continued). Take notice of the transition $i-1=6 \rightarrow i=7$ in these trellises. From
\begin{displaymath}
(N_6 \vert \mbox{\boldmath $\bar g$}_7 \vert N_7)=\left(
\begin{array}{ccc|c|ccc}
1 & 0 & 1 & 1 & 0 & 0 & 0 \\
0 & 0 & 0 & 0 & 0 & 0 & 0 \\
0 & 0 & 0 & 0 & 0 & 0 & 0 \\
0 & 0 & 0 & 1 & 1 & 0 & 1
\end{array}
\right) ,
\end{displaymath}
we have
\begin{eqnarray}
\mbox{ker}(N_6) &=& \{0000, 0001, 0010, 0011, \nonumber \\
&& 0100, 0101, 0110, 0111\} \nonumber \\
\mbox{ker}(N_7) &=& \{0000, 0010, 0100, 0110, \nonumber \\
&& 1000, 1010, 1100, 1110\} \nonumber \\
\mbox{ker}(N_6, \mbox{\boldmath $\bar g$}_7, N_7) &=& \{0000, 0010, 0100, 0110\} . \nonumber
\end{eqnarray}
Then
\begin{eqnarray}
\mbox{dim ker}(N_6) &=& 3 \nonumber \\
\mbox{dim ker}(N_7) &=& 3 \nonumber \\
\mbox{dim ker}(N_6, \mbox{\boldmath $\bar g$}_7, N_7) &=& 2 . \nonumber
\end{eqnarray}
Similarly, from
\begin{eqnarray}
\lefteqn{(\hat N_6 \vert \mbox{\boldmath $h$}_7 \vert \hat N_7)=(N_6^T \vert \mbox{\boldmath $h$}_7 \vert N_7^T)} \nonumber \\
&& =\left(
\begin{array}{cccc|c|cccc}
1 & 0 & 0 & 0 & 1 & 0 & 0 & 0 & 1 \\
0 & 0 & 0 & 0 & 0 & 0 & 0 & 0 & 0 \\
1 & 0 & 0 & 0 & 1 & 0 & 0 & 0 & 1
\end{array}
\right) , \nonumber
\end{eqnarray}
we have
\begin{eqnarray}
\mbox{ker}(\hat N_6) &=& \{000, 010, 101, 111\} \nonumber \\
\mbox{ker} (\hat N_7) &=& \{000, 010, 101, 111\} \nonumber \\
\mbox{ker} (\hat N_6, \mbox{\boldmath $h$}_7, \hat N_7) &=& \{000, 010, 101, 111\} . \nonumber
\end{eqnarray}
Then
\begin{eqnarray}
\mbox{dim ker}(\hat N_6) &=& 2 \nonumber \\
\mbox{dim ker}(\hat N_7) &=& 2 \nonumber \\
\mbox{dim ker}(\hat N_6, \mbox{\boldmath $h$}_7, \hat N_7) &=& 2 . \nonumber
\end{eqnarray}
Hence, we have
\begin{displaymath}
k-\mbox{dim ker}(N_{i-1}, \mbox{\boldmath $\bar g$}_i, N_i)=4-2=2
\end{displaymath}
\begin{eqnarray}
\lefteqn{(n-k+1)+\mbox{dim ker}(\hat N_{i-1}, \mbox{\boldmath $h$}_i, \hat N_i)} \nonumber \\
&& \quad -\mbox{dim ker}(\hat N_{i-1})-\mbox{dim ker}(\hat N_i) \nonumber \\
&& =(7-4+1)+2-2-2=2 . \nonumber
\end{eqnarray}
The second equality is confirmed in a similar way.

% end of file

%% file: sec.5-c.tex
\section{Constructing a KV Trellis From an Extended Minimal-Span Generator Matrix}
A conventional BCJR trellis can be constructed using a minimal-span generator matrix (MSGM)~\cite{mc 96}. In this section, we show that this construction is extended to tail-biting trellises. First, we show an example and then extend to a general case.
\subsection{An example}
Consider the generator matrix
\begin{displaymath}
G=\left(
\begin{array}{ccccccc}
0& 0 & 0 & \mbox{\boldmath $1$} & \mbox{\boldmath $1$} & \mbox{\boldmath $0$} & \mbox{\boldmath $1$} \\
\mbox{\boldmath $1$}& \mbox{\boldmath $1$} & \mbox{\boldmath $0$} & \mbox{\boldmath $1$} & 0 & 0 & 0 \\
0& 0 & \mbox{\boldmath $1$} & \mbox{\boldmath $1$} & \mbox{\boldmath $0$} & \mbox{\boldmath $1$} & 0 \\
\mbox{\boldmath $1$}& \mbox{\boldmath $0$} & \mbox{\boldmath $1$} & 0 & 0 & 0 & \mbox{\boldmath $1$}
\end{array}
\right)
\begin{array}{l}
$[4, 7]$ \\
$[1, 4]$ \\
$[3, 6]$ \\
$[7, 3]$ .
\end{array}
\end{displaymath}
\par
{\it Remark 1:} This generator matrix has been considered in Section III-B (Example 1) and Section III-D (Example 2). We also remark that the numbering of indices for a codeword is shifted by $1$ compared to that in~\cite{koe 03} (also in~\cite{glu 111}).
\par
Note that left/right end-points of the spans of $G$ are distinct. Hence, $G$ can be regarded as a kind of MSGM~\cite{mc 96}. We call such an MSGM an {\it extended minimal-span generator matrix} (e-MSGM). According to McEliece~\cite{mc 96}, let us define $A_i$ and $B_i$ as follows:
\begin{eqnarray}
A_i &=& \{j:~i \in [a_j, b_j]\},~\mbox{for}~i=1, \cdots, n \\
A_0 &=& A_n \\
B_i &=& A_i \cap A_{i+1},~\mbox{for}~i=0, \cdots, n-1 \\
B_n &=& B_0 ,
\end{eqnarray}
where the periodicity of a tail-biting trellis is taking into account. Here we assume an additional condition:
\par
$(\sharp)$ Let $[a_l, b_l]$ be the circular span of $\mbox{\boldmath $g$}_l$ such that $a_l=i+1$ and $b_l=i$. In this case, $l$ is not contained in $B_i$.
\par
We denote the cardinalities of $A_i$ and $B_i$ by $\alpha_i$ and $\beta_i$, respectively. Table I gives the $A_i\mbox{'s}$, $B_i\mbox{'s}$, $\alpha_i\mbox{'s}$, and $\beta_i\mbox{'s}$.
\begin{table}[htb]
\caption{$A_i$'s, $B_i$'s, $\alpha_i$'s, $\beta_i$'s}
\label{table:1}
\begin{center}
\begin{tabular}{c*{4}{|c}}
$i$ & $A_i$ & $B_i$ & $\alpha_i$ & $\beta_i$ \\
\hline
0 & \{1, 4\} & \{4\} & 2 & 1 \\ 
1 & \{2, 4\} & \{2, 4\} & 2 & 2 \\
2 & \{2, 4\} & \{2, 4\} & 2 & 2 \\ 
3 & \{2, 3, 4\} & \{2, 3\} & 3 & 2 \\
4 & \{1, 2, 3\} & \{1, 3\} & 3 & 2 \\ 
5 & \{1, 3\} & \{1, 3\} & 2 & 2 \\
6 & \{1, 3\} & \{1\} & 2 & 1 \\ 
7 & \{1, 4\} & \{4\} & 2 & 1 
\end{tabular}
\end{center}
\end{table}
Based on this table, we can construct a tail-biting trellis using the method in~\cite{mc 96}. Let $u$ a binary $\alpha_i$-tuple and define the variables $\mbox{init}(u)$, $\mbox{fin}(u)$, and $\lambda(u)$ as follows:
\begin{eqnarray}
\mbox{init}(u) &=& u \cap B_{i-1} \\
\mbox{fin}(u) &=& u \cap B_i \\
\lambda(u) &=& u\cdot \mbox{\boldmath $\bar g$}_i' ,
\end{eqnarray}
where $\mbox{\boldmath $\bar g$}_i'=\mbox{\boldmath $\bar g$}_i \cap A_i$. The notation ``$u \cap B$'' represents the binary vector obtained by extracting the components of $u$ corresponding to the elements of $B$. Then for edge spaces $E_i~(1 \leq i \leq 7)$, we have the following tables. A tail-biting trellis is constructed based on these tables. The resulting tail-biting trellis is shown in Fig.10.
\begin{center}
\begin{tabular}{c*{3}{|c}}
\multicolumn {4}{c}{$E_1$} \\
$u$ & $\mbox{init}(u)$ & $\mbox{fin}(u)$ & $\lambda(u)$ \\
\hline
24 & 4 & 24 & [11] \\ 
00 & 0 & 00 & 0 \\
01 & 1 & 01 & 1 \\ 
10 & 0 & 10 & 1 \\
11 & 1 & 11 & 0 
\end{tabular}
\end{center}
\begin{center}
\begin{tabular}{c*{3}{|c}}
\multicolumn {4}{c}{$E_2$} \\
$u$ & $\mbox{init}(u)$ & $\mbox{fin}(u)$ & $\lambda(u)$ \\
\hline
24 & 24 & 24 & [10] \\ 
00 & 00 & 00 & 0 \\
01 & 01 & 01 & 0 \\ 
10 & 10 & 10 & 1 \\
11 & 11 & 11 & 1 
\end{tabular}
\end{center}
\begin{center}
\begin{tabular}{c*{3}{|c}}
\multicolumn {4}{c}{$E_3$} \\
$u$ & $\mbox{init}(u)$ & $\mbox{fin}(u)$ & $\lambda(u)$ \\
\hline
234 & 24 & 23 & [011] \\ 
000 & 00 & 00 & 0 \\
001 & 01 & 00 & 1 \\ 
010 & 00 & 01 & 1 \\
011 & 01 & 01 & 0 \\
100 & 10 & 10 & 0 \\
101 & 11 & 10 & 1 \\ 
110 & 10 & 11 & 1 \\
111 & 11 & 11 & 0 
\end{tabular}
\end{center}
\begin{center}
\begin{tabular}{c*{3}{|c}}
\multicolumn {4}{c}{$E_4$} \\
$u$ & $\mbox{init}(u)$ & $\mbox{fin}(u)$ & $\lambda(u)$ \\
\hline
123 & 23 & 13 & [111] \\ 
000 & 00 & 00 & 0 \\
001 & 01 & 01 & 1 \\ 
010 & 10 & 00 & 1 \\
011 & 11 & 01 & 0 \\
100 & 00 & 10 & 1 \\
101 & 01 & 11 & 0 \\ 
110 & 10 & 10 & 0 \\
111 & 11 & 11 & 1 
\end{tabular}
\end{center}
\begin{center}
\begin{tabular}{c*{3}{|c}}
\multicolumn {4}{c}{$E_5$} \\
$u$ & $\mbox{init}(u)$ & $\mbox{fin}(u)$ & $\lambda(u)$ \\
\hline
13 & 13 & 13 & [10] \\ 
00 & 00 & 00 & 0 \\
01 & 01 & 01 & 0 \\ 
10 & 10 & 10 & 1 \\
11 & 11 & 11 & 1 
\end{tabular}
\end{center}
\begin{center}
\begin{tabular}{c*{3}{|c}}
\multicolumn {4}{c}{$E_6$} \\
$u$ & $\mbox{init}(u)$ & $\mbox{fin}(u)$ & $\lambda(u)$ \\
\hline
13 & 13 & 1 & [01] \\ 
00 & 00 & 0 & 0 \\
01 & 01 & 0 & 1 \\ 
10 & 10 & 1 & 0 \\
11 & 11 & 1 & 1 
\end{tabular}
\end{center}
\begin{center}
\begin{tabular}{c*{3}{|c}}
\multicolumn {4}{c}{$E_7$} \\
$u$ & $\mbox{init}(u)$ & $\mbox{fin}(u)$ & $\lambda(u)$ \\
\hline
14 & 1 & 4 & [11] \\ 
00 & 0 & 0 & 0 \\
01 & 0 & 1 & 1 \\ 
10 & 1 & 0 & 1 \\
11 & 1 & 1 & 0 
\end{tabular}
\end{center}
\begin{figure}[tb]
\begin{center}
\includegraphics[width=8.0cm,clip]{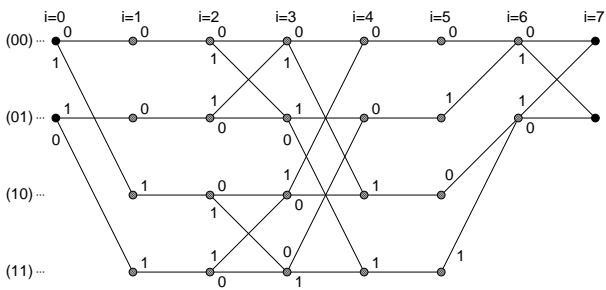}
\end{center}
\caption{The tail-biting trellis constructed based on the e-MSGM.}
\label{fig:10}
\end{figure}

\subsection{Generalization}
As is shown in~\cite{glu 111} and~\cite{glu 112}, the state matrix $M_i$ of a KV-trellis is given as follows:
\begin{eqnarray}
M_i &=& \left(
\begin{array}{ccc}
\mu_i^1 & &  \\
 & \ddots & \\
 & & \mu_i^k
\end{array}
\right) \nonumber \\
\mu_i^l &=& \left\{
\begin{array}{ll}
1, & \mbox{if}~i \in (a_l, b_l], \\
0, & \mbox{if}~i \notin (a_l, b_l] .
\end{array} \right. \nonumber
\end{eqnarray}
\par
{\it Remark 2:} In~\cite{koe 03}, the span of a generator $\mbox{\boldmath $g$}_l$, denoted by $[\mbox{\boldmath $g$}_l]$, is defined as a semiopen interval $(a_l, b_l]$ such that the corresponding closed interval $[a_l, b_l]$ contains all the nonzero positions of $\mbox{\boldmath $g$}_l$. Now, consider the condition $\mu_i^l=1$. This means that $i-1 \in [a_l, b_l]$ and $i \in [a_l, b_l]$. Here take notice of the numbering of indices for a codeword in the two papers, i.e., one is by Koetter and Vardy (also by Gluesing-Luerssen and Weaver) and the other is by McEliece (also by Nori and Shankar). We see that the numbering of indices is shifted by $1$ between them. Hence, in the notation of McEliece, the condition $\mu_i^l=1$ is equivalent to $i \in [a_l, b_l]$ and $i+1 \in [a_l, b_l]$ (i.e., $\mbox{\boldmath $g$}_l$ is ``active'' at levels $i$ and $i+1$). Taking these facts into consideration, we follow McEliece in this section.
\par
Thus the next lemma has been proved.
\begin{lem}
\begin{equation}
\mu_i^l=1 \leftrightarrow l \in B_i .
\end{equation}
\end{lem}
\par
Here consider the product $u\cdot M_i$. When we compute
\begin{displaymath}
(u_1, \cdots, u_l, \cdots, u_k)\left(
\begin{array}{ccccc}
\mu_i^1 & & & & \\
 & \ddots & & & \\
 & & \mu_i^l & & \\
 & & & \ddots & \\
 & & & & \mu_i^k
\end{array}
\right) ,
\end{displaymath}
the product has the form $(\cdots, u_l, \cdots)$ if $\mu_i^l=1$. On the other hand, if $l \in B_i$, then $u_l$ is extracted in the operation of $u \cap B_i$. Hence, it follows that
\begin{eqnarray}
u_l \in u \cap B_i &\leftrightarrow& l \in B_i \nonumber \\
&\leftrightarrow& \mu_i^l=1 \nonumber \\
&\leftrightarrow& (\cdots, u_l, \cdots) \in \mbox{im}M_i=V_i .
\end{eqnarray}
Accordingly, we have
\begin{itemize}
\item $u_l \in \mbox{init}(u) \leftrightarrow (\cdots, u_l, \cdots) \in V_{i-1}$,
\item $u_l \in \mbox{fin}(u) \leftrightarrow (\cdots, u_l, \cdots) \in V_i$,
\item $\lambda(u)=u \cdot \mbox{\boldmath $\bar g$}_i=c_i$.
\end{itemize}
Here we need not consider all the components of $u$. From the definition, $B_{i-1} \subset A_i$ and $B_i \subset A_i$ hold. Thus we can restrict $u$ to $u \cap A_i$ (i.e., $u$ is a binary $\alpha_i$-tuple). Hence, in the calculation of $\lambda(u)$, $\mbox{\boldmath $\bar g$}_i$ can be replaced by $\mbox{\boldmath $\bar g$}_i'=\mbox{\boldmath $\bar g$}_i \cap A_i$. We see that these are equivalent to the tables for $E_i$.
\par
Thus we have shown the following.
\begin{pro}
Consider a KV trellis $T_{G, S}$. Denote by $T_{[G, S]}$ the corresponding tail-biting trellis constructed by regarding $(G, S)$ as an e-MSGM. Then $T_{[G, S]}$ is identical to $T_{G, S}$.
\end{pro}
\par
In fact, we observe that the tail-biting trellis in Fig.10 is identical to the KV trellis in Fig.8.
\par
Moreover, we have the following.
\begin{pro}
Suppose that a KV trellis $T_{G, S}$ is constructed using the proposed algebraic method. Then the resulting tail-biting trellis $T_{alg}$ is isomorphic to $T_{[G, S]}$. If the coset representatives (in terms of {\boldmath $u$}) are placed in ascending order at each level $i$, then $T_{alg}$ is identical to $T_{[G, S]}$.
\end{pro}
\begin{IEEEproof}
For the proposed algebraic construction,
\begin{displaymath}
E_i \cong \mbox{F}^k/\mbox{ker}(M_{i-1}, \mbox{\boldmath $\bar g$}_i, M_i),~\mbox{for}~i=1, \cdots, n 
\end{displaymath}
holds. For $\mbox{\boldmath $u$},~\mbox{\boldmath $u$}' \in \mbox{F}^k$, let $\mbox{\boldmath $c$}$ and $\mbox{\boldmath $c$}'$ be the corresponding codewords. Also, suppose that $\mbox{\boldmath $u$}-\mbox{\boldmath $u$}' \in \mbox{ker}(M_{i-1}, \mbox{\boldmath $\bar g$}_i, M_i)$. This means the following (cf. Proposition 2):
\begin{itemize}
\item At level $i-1$, $u_l=u_l'$ holds for $l$ such that $\mu_{i-1}^l=1$,
\item At level $i$, $u_l=u_l'$ holds for $l$ such that $\mu_i^l=1$,
\item $c_i=c_i'$.
\end{itemize}
The above are further rephrased as follows:
\begin{itemize}
\item If $u_l \in \mbox{init}(u), u_l' \in \mbox{init}(u')$, then $u_l=u_l'$,
\item If $u_l \in \mbox{fin}(u), u_l' \in \mbox{fin}(u')$, then $u_l=u_l'$,
\item $c_i=c_i'$.
\end{itemize}
\end{IEEEproof}
\par
Finally, we remark that if a generator matrix $G$ has a form of e-MSGM, then the corresponding tail-biting BCJR trellis $T_{(G, H, \Theta)}$ can also be constructed according to the procedure described above. In fact, it has been shown that the trellises $T_{G, S}$ and $T_{(G, H, \Theta)}$ are isomorphic~\cite[Theorem IV.11]{glu 111}.

%% file: sec.app.tex
Consider an $(n, k)$ linear block code $C$. Let $i$ be any level. Forney~\cite{forn 88} defined the past and future subcodes $P_i$ and $F_i$ as follows (cf.~\cite{mc 96}):
\begin{eqnarray}
P_i &=& \{\mbox{\boldmath $c$} \in C: c_{i+1}=c_{i+2}=\cdots =c_n=0\} \\
F_i &=& \{\mbox{\boldmath $c$} \in C: c_1=c_2=\cdots =c_i=0\} .
\end{eqnarray}
We denote their dimensions by $p_i$ and $f_i$, respectively:
\begin{eqnarray}
p_i &=& \mbox{dim}P_i,~i=0, \cdots, n-1 \\
f_i &=& \mbox{dim}F_i,~i=1, \cdots, n .
\end{eqnarray}
Consider the associated conventional trellis (i.e., the Forney trellis). Then for the state and edge spaces $V_i$ and $E_i$ at level $i$, we have
\begin{eqnarray}
\vert V_i \vert &=& 2^{k-p_i-f_i} \\
\vert E_i \vert &=& 2^{k-p_{i-1}-f_i} .
\end{eqnarray}
A similar result holds for the dual code $C^{\perp}$ of $C$. Let $\hat p_i$ and $\hat f_i$ be the corresponding variables. Then it follows that
\begin{eqnarray}
\hat p_i &=& f_i+i-k \\
\hat f_i &=& p_i-i+(n-k) .
\end{eqnarray}
\par
On the other hand, as is shown in~\cite{mc 96}, the same conventional trellis (i.e., the BCJR trellis) is constructed based on the MSGM. In this case, we have
\begin{eqnarray}
\vert V_i \vert &=& 2^{\beta_i} \\
\vert E_i \vert &=& 2^{\alpha_i} ,
\end{eqnarray}
where $\alpha_i=\vert A_i \vert $ and $\beta_i=\vert B_i \vert $ (see Section V). Then it follows that
\begin{eqnarray}
\beta_i &=& k-p_i-f_i \\
\alpha_i &=& k-p_{i-1}-f_i .
\end{eqnarray}
Similarly, we have
\begin{eqnarray}
\hat \beta_i &=& (n-k)-\hat p_i-\hat f_i \\
\hat \alpha_i &=& (n-k)-\hat p_{i-1}-\hat f_i .
\end{eqnarray}
Based on these equations, we first derive the relations between $(\alpha_i, \beta_i)$ and $(\hat \alpha_i, \hat \beta_i)$ for conventional trellises. At this point, the derived relations hold only for conventional trellises. However, note that the relations are expressed in terms of $\alpha_i$, $\beta_i$, $\hat \alpha_i$, and $\hat \beta_i$. That is, variables such as $p_i$ and $f_i$ are not contained in the relations. On the other hand, we already have shown that the trellis construction based on MSGM's is extended to the construction based on e-MSGM's. This implies that the relations between $(\alpha_i, \beta_i)$ and $(\hat \alpha_i, \hat \beta_i)$ still hold for ``tail-biting trellises''. This is a basic idea for the proof.
\par
Now we go back to the proof. First, note that
\begin{displaymath}
\alpha_i=k-p_{i-1}-f_i .
\end{displaymath}
From the duality formulae, it follows that
\begin{eqnarray}
\hat p_i &=& f_i+i-k \nonumber \\
\hat f_{i-1} &=& p_{i-1}-(i-1)+(n-k) . \nonumber
\end{eqnarray}
Adding each side, we have
\begin{displaymath}
\hat p_i+\hat f_{i-1}=(p_{i-1}+f_i)+1+n-2k .
\end{displaymath}
Hence,
\begin{displaymath}
\alpha_i+(\hat p_i+\hat f_{i-1})=1+(n-k)
\end{displaymath}
holds. On the other hand, from
\begin{eqnarray}
\hat \beta_{i-1} &=& (n-k)-\hat p_{i-1}-\hat f_{i-1} \nonumber \\
\hat \alpha_i &=& (n-k)-\hat p_{i-1}-\hat f_i , \nonumber
\end{eqnarray}
it follows that
\begin{displaymath}
\hat \beta_{i-1}-\hat \alpha_i=-\hat f_{i-1}+\hat f_i .
\end{displaymath}
By combining this with
\begin{displaymath}
\hat \beta_i=(n-k)-\hat p_i-\hat f_i ,
\end{displaymath}
we have
\begin{displaymath}
-\hat \alpha_i+\hat \beta_{i-1}+\hat \beta_i=(n-k)-(\hat p_i+\hat f_{i-1}) .
\end{displaymath}
Moreover, by combining this with
\begin{displaymath}
\alpha_i+(\hat p_i+\hat f_{i-1})=1+(n-k) ,
\end{displaymath}
we finally have
\begin{displaymath}
\alpha_i=-\hat \alpha_i+\hat \beta_{i-1}+\hat \beta_i+1 .
\end{displaymath}
The equality
\begin{displaymath}
\hat \alpha_i=-\alpha_i+\beta_{i-1}+\beta_i+1
\end{displaymath}
is derived in a similar way.

% end of file

%% file: alg.t-d.bbl
\begin{thebibliography}{99}
%\bibitem{ander 84}J.~B.~Anderson and S.~Mohan, ``Sequential coding algorithms: A survey and cost analysis,'' {\em IEEE Trans. Commun.}, vol.~COM-32, no.~2, pp.~169--176, Feb. 1984.
%\bibitem{arie 95}M.~Ariel and J.~Snyders, ``Soft syndrome decoding of binary convolutional codes,'' {\em IEEE Trans. Commun.}, vol.~43, no.~2/3/4, pp.~288--297, Feb./March/April 1995.
\bibitem{arie 981}M.~Ariel and J.~Snyders, ``Minimal error-trellises for linear block codes,'' in {\em Proc. ISIT1998}, p.~228, Aug. 1998.
\bibitem{arie 982}\bysame, ``Error-trellises for convolutional codes--Part I: Construction,'' {\em IEEE Trans.~Commun.}, vol.~46, no.~12, pp.~1592--1601, Dec. 1998.
%\bibitem{arie 99}\bysame, ``Error-trellises for convolutional codes--Part II: Decoding methods,'' {\em IEEE Trans.~Commun.}, vol.~47, no.~7, pp.~1015--1024, July 1999.
%\bibitem{arik 08}E.~Arikan, ``Channel polarization: A method for constructing capacity-achieving codes,'' in {\em Proc. ISIT2008}, pp.~1173--1177, July 2008.
\bibitem{bahl 74}L.~R.~Bahl, J.~Cocke, F.~Jelinek, and J.~Raviv, ``Optimal decoding of linear codes for minimizing symbol error rate,'' {\em IEEE Trans.~Inform.~Theory}, vol.~IT-20, no.~2, pp.~284--287, March 1974.
%\bibitem{bene 98}S.~Benedetto, D.~Divsalar, and J.~Hagenauer, Eds., Special Issue on ``Concatenated coding techniques and iterative decoding: Sailing toward channel capacity,'' {\em IEEE J. Select. Areas Commun.}, vol.~16, no.~2, Feb. 1998.
\bibitem{con 15}D.~Conti and N.~Boston, ``On the algebraic structure of linear tail-biting trellises,'' {\em IEEE Trans.~Inform.~Theory}, vol.~61, no.~5, pp.~2283--2299, May 2015.
%\bibitem{cos 09}D.~Costello, S.~Lin, T.~Richardson, W.~Ryan, R.~Urbanke, and R.~Wesel, Eds., Special Issue on ``Capacity approaching codes,'' {\em IEEE J. Select. Areas Commun.}, vol.~27, no.~6, Aug. 2009.
%\bibitem{dhol 94}A.~Dholakia, {\em Introduction to Convolutional Codes with Applications}. Norwell, MA: Kluwer, 1994.
%\bibitem{eli 55}P.~Elias, ``Coding for noisy channels,'' {\em I.R.E. Convention Record}, Part IV, pp.~37--46, 1955.
%\bibitem{fano 63}R.~M.~Fano, ``A heuristic discussion of probabilistic decoding,'' {\em IEEE Trans.~Inform.~Theory}, vol.~IT-9, no.~2, pp.~64--74, April 1963.
%\bibitem{fels 99}A.~J.~Felstroem and K.~ Sh.~Zigangirov, ``Time-varying periodic convolutional codes with low-density parity-check matrix,'' {\em IEEE Trans.~Inform.~Theory}, vol.~45, no.~6, pp.~2181--2191, Sep. 1999.
\bibitem{forn 70}G.~D.~Forney, Jr., ``Convolutional codes I: Algebraic structure,'' {\em IEEE Trans.~Inform.~Theory}, vol.~IT-16, no.~6, pp.~720--738, Nov. 1970.
%\bibitem{forn 731}\bysame, ``The Viterbi algorithm,'' {\em Proc. of the IEEE}, vol.~61, no.~3, pp.~268--278, March 1973.
\bibitem{forn 732}\bysame, ``Structural analysis of convolutional codes via dual codes,'' {\em IEEE Trans.~Inform.~Theory}, vol.~IT-19, no.~4, pp.~512--518, July 1973.
%\bibitem{forn 74}\bysame, ``Convolutional codes II. Maximum-likelihood decoding,'' {\em Information and Control}, vol.~25, no.~3, pp.~222--266, July 1974.
%\bibitem{forn 75}\bysame, ``Minimal bases of rational vector spaces, with applications to multivariable linear systems,'' {\em SIAM J. on Control}, vol.~13, no.~3, pp.~493--520, May 1975.
\bibitem{forn 88}\bysame, ``Coset codes--Part II: Binary lattices and related codes,'' {\em IEEE Trans.~Inform.~Theory}, vol.~34, no.~5, pp.~1152--1187 (Appendix A), Sept. 1988.
\bibitem{forn 94}\bysame, ``Dimension/length profiles and trellis complexity of linear block codes,'' {\em IEEE Trans.~Inform.~Theory}, vol.~40, no.~6, pp.~1741--1752, Nov. 1994.
%\bibitem{galla 63}R.~G.~Gallager, {\em Low-Density Parity-Check Codes}. Cambridge, MA: MIT Press, 1963.
\bibitem{glu 111}H.~Gluesing-Luerssen and E. A. Weaver, ``Linear tail-biting trellises: Characteristic generators and the BCJR-construction,'' {\em IEEE Trans.~Inform.~Theory}, vol.~57, no.~2, pp.~738--751, Feb. 2011.
\bibitem{glu 112}\bysame, ``Characteristic generators and dualization for tail-biting trellises,'' {\em IEEE Trans.~Inform.~Theory}, vol.~57, no.~11, pp.~7418--7430, Nov. 2011.
\bibitem{glu 13}H.~Gluesing-Luerssen and G.~D.~Forney, Jr., ``Local irreducibility of tail-biting trellises,'' {\em IEEE Trans.~Inform.~Theory}, vol.~59, no.~10, pp.~6597--6610, Oct. 2013.
\bibitem{hart 76}C.~R.~P.~Hartmann and L.~D.~Rudolph, ``An optimum symbol-by-symbol decoding rule for linear codes,'' {\em IEEE Trans.~Inform.~Theory}, vol.~IT-22, no.~5, pp.~514--517, Sept. 1976.
\bibitem{joha 99}R.~Johannesson and K.~S.~Zigangirov, {\em Fundamentals of Convolutional Coding}. New York: IEEE Press, 1999.
%\bibitem{kasa 11}Š}ˆäŒ''¾, ``[µ'ҍu‰‰]‹óŠÔŒ‹‡•"†'Æ'»'̉ž—p,'' "dŽqî•ñ'ʐMŠw‰ï‹ZpŒ¤‹†•ñ, IT2011-24, pp.~1--8, Sept. 2011.
%\bibitem{kubo 85}‹v•Û"cŽüŽ¡CŒS•Ž¡C‰Á"¡CŽO, ``SST (Scarce State Transition)Œ^ƒrƒ^ƒr•œ†‰ñ˜H,''"dŽq'ʐMŠw‰ï˜_•¶Ž, vol.~J68-B, no.~1, pp.~38--45, Jan. 1985.
%\bibitem{kai 681}T.~Kailath, ``An innovations approach to least-squares estimation--Part I: Linear filtering in additive white noise,'' {\em IEEE Trans. Automatic Control}, vol.~AC-13, no.~6, pp.~646--655, Dec. 1968.
%\bibitem{kai 682}T.~Kailath and P.~Frost, ``An innovations approach to least-squares estimation--Part II: Linear smoothing in additive white noise,'' {\em IEEE Trans. Automatic Control}, vol.~AC-13, no.~6, pp.~655--660, Dec. 1968.
\bibitem{kie 96}A.~B.~Kiely, S.~J.~Dolinar, Jr., R.~J.~McEliece, L.~L.~Ekroot, and W.~Lin, ``Trellis decoding complexity of linear block codes,'' {\em IEEE Trans.~Inform.~Theory}, vol.~42, no.~6, pp.~1687--1697, Nov. 1996.
\bibitem{koe 03}R. Koetter and A. Vardy, ``The structure of tail-biting trellises: Minimality and basic principles,'' {\em IEEE Trans.~Inform.~Theory}, vol.~49, no.~9, pp.~2081--2105, Sep. 2003.
\bibitem{ks 95}F.~R.~Kschischang and V.~Sorokine, ``On the trellis structure of block codes,'' {\em IEEE Trans.~Inform.~Theory}, vol.~41, no.~6, pp.~1924--1937, Nov. 1995.
\bibitem{lin 98}S.~Lin, T.~Kasami, T.~Fujiwara, and M.~Fossorier, {\em Trellises and Trellis-Based Decoding Algorithms for Linear Block Codes}. Norwell, MA: Kluwer, 1998.
\bibitem{lin 00}S.~Lin and R.~Y.~Shao, ``General structure and construction of tail biting trellises for linear block codes,'' in {\em Proc. ISIT2000}, p.~117, June 2000.
%\bibitem{mori 10}X—§•½, ``[µ'ҍu‰‰]Polar•"†'ɂ'¢'Ä,'' "dŽqî•ñ'ʐMŠw‰ï‹ZpŒ¤‹†•ñ, IT2010-41, pp.~43--49, Sept. 2010.
\bibitem{ma 86}H. H. Ma and J. K. Wolf, ``On tail biting convolutional codes,'' {\em IEEE Trans. Commun.}, vol.~COM-34, no.~2, pp.~104--111, Feb. 1986.
\bibitem{mc 96}R.~J.~McEliece, ``On the BCJR trellis for linear block codes,'' {\em IEEE Trans.~Inform.~Theory}, vol.~42, no.~4, pp.~1072--1092, July 1996.
\bibitem{mc 962}R.~J.~McEliece and W.~Lin, ``The trellis complexity of convolutional codes,'' {\em IEEE Trans.~Inform.~Theory}, vol.~42, no.~6, pp.~1855--1864, Nov. 1996.
\bibitem{mud 88}D.~J.~Muder, ``Minimal trellises for block codes,'' {\em IEEE Trans.~Inform.~Theory}, vol.~34, no.~5, pp.~1049--1053, Sept. 1988.
\bibitem{nori 06}A. V. Nori and P. Shankar, ``Unifying views of tail-biting trellis constructions for linear block codes,'' {\em IEEE Trans.~Inform.~Theory}, vol.~52, no.~10, pp.~4431--4443, Oct. 2006.
\bibitem{pire 88}P.~Piret, {\em Convolutional Codes--An Algebraic Approach}. Cambridge, MA: MIT Press, 1988.
%\bibitem{ried 98}S.~Riedel, ``Symbol-by-symbol MAP decoding algorithm for high-rate convolutional codes that use reciprocal dual codes,'' {\em IEEE J.~Select.~Areas Commun.}, vol.~16, no.~2, pp.~175--185, Feb. 1998.
%\bibitem{scha 75}J.~P.~M.~Schalkwijk and A.~J.~Vinck, ``Syndrome decoding of convolutional codes,'' {\em IEEE Trans. Commun.}, vol.~COM-23, pp.~789--792, July 1975.
%\bibitem{scha 76}\bysame, ``Syndrome decoding of binary rate-$1/2$ convolutional codes,'' {\em IEEE Trans.~Commun.}, vol.~COM-24, no.~9, pp.~977--985, Sept. 1976.
%\bibitem{scha 78}J.~P.~M.~Schalkwijk, A.~J.~Vinck, and K.~A.~Post, ``Syndrome decoding of binary rate-$k/n$ convolutional codes,'' {\em IEEE Trans.~Inform.~Theory}, vol.~IT-24, no.~5, pp.~553--562, Sept. 1978.
\bibitem{sha 00}Y.~Shany and Y.~Be'ery, ``Linear tail-biting trellises, the square-root bound, and applications for Reed-Muller codes,'' {\em IEEE Trans.~Inform.~Theory}, vol.~46, no.~4, pp.~1514--1523, July 2000.
%\bibitem{sho 06}A.~Shokrollahi, ``Raptor codes,'' {\em IEEE Trans.~Inform.~Theory}, vol.~52, no.~6, pp.~2551--2567, June 2006.
\bibitem{sido 94}V.~Sidorenko and V.~Zyablov, ``Decoding of convolutional codes using a syndrome trellis,'' {\em IEEE Trans.~Inform.~Theory}, vol.~40, no.~5, pp.~1663--1666, Sept. 1994.
%\bibitem{sie 011}P.~H.~Siegel, D.~Divsalar, E.~Eleftheriou, J.~Hagenauer, and D.~Rowitch, Eds., Special Issue on ``The turbo principle: From theory to practice I,'' {\em IEEE J. Select. Areas Commun.}, vol.~19, no.~5, May 2001.
%\bibitem{sie 012}P.~H.~Siegel, D.~Divsalar, E.~Eleftheriou, J.~Hagenauer, and D.~Rowitch, Eds., Special Issue on ``The turbo principle: From theory to practice II,'' {\em IEEE J. Select. Areas Commun.}, vol.~19, no.~9, Sept. 2001.
%\bibitem{taji 88}M.~Tajima and H.~Suzuki, ``On the general equivalence between the SST-type Viterbi decoder and the syndrome decoder for convolutional codes,'' in {\em Abstracts of ISIT1988}, pp.51-52, June 1988.
%\bibitem{taji 96}M.~Tajima, ``On the structure of an SST Viterbi decoder for general rate $(n-1)/n$ convolutional codes viewed in the light of syndrome decoding,'' {\em IEICE Trans. Fundamentals}, vol.~E79-A, no.~9, pp.~1447--1449, Sep. 1996.
\bibitem{taji 02}M.~Tajima, K.~Shibata, and Z.~Kawasaki, ``Modification of syndrome trellises for high-rate convolutional codes,'' {\em IEICE Technical Report}, IT2002-2, pp.~7--12, May 2002.
%\bibitem{taji 031}\bysame, ``On the equivalence between Scarce-State-Transition Viterbi decoding and syndrome decoding of convolutional codes,'' in {\em Proc. ISIT2003}, p.~304, July 2003.
%\bibitem{taji 032}\bysame, ``On the equivalence between Scarce-State-Transition Viterbi decoding and syndrome decoding of convolutional codes,'' {\em IEICE Trans. Fundamentals}, vol.~E86-A, no.~8, pp.~2107--2116, Aug. 2003.
%\bibitem{taji 033}\bysame, ``Relation between encoder and syndrome former variables and symbol reliability estimation using a syndrome trellis,'' {\em IEEE Trans. Commun.}, vol.~51, no.~9, pp.~1474--1484, Sept. 2003.
%\bibitem{taji 04}M.~Tajima, K.~Okino, and T.~Miyagoshi, ``Trellis complexity analysis of convolutional codes using reciprocal dual codes,'' in {\em Proc. SITA2004 (Japanese Ed.)}, pp.~91--94, Dec. 2004.
%\bibitem{taji 06}\bysame, ``State-complexity reduction for convolutional codes using trellis-module integration,'' {\em IEICE Trans. Fundamentals}, vol.~E89-A, no.~10, pp.~2466--2474, Oct. 2006.
\bibitem{taji 07}M.~Tajima, K.~Okino, and T.~Miyagoshi, ``Minimal code(error)-trellis module construction for rate-$k/n$ convolutional codes: Extension of Yamada-Harashima-Miyakawa's construction,'' {\em IEICE Trans. Fundamentals}, vol.~E90-A, no.~11, pp.~2629--2634, Nov. 2007.
%\bibitem{taji 09}\bysame, ``Error-trellis construction for convolutional codes using shifted error/syndrome-subsequences,'' {\em IEICE Trans. Fundamentals}, vol.~E92-A, no.~8, pp.~2086--2096, Aug. 2009.
%\bibitem{taji 10}\bysame, ``Error-trellis state complexity of LDPC convolutional codes based on circulant matrices,'' in {\em Proc. ISITA2010}, pp.~19--24, Oct. 2010.
%\bibitem{taji 111}\bysame, ``[Invited Talk] Code/error-trellis construction for convolutional codes using shifted code/error-subsequences,'' {\em IEICE Technical Report}, IT2010-73, pp.~37--44, March 2011.
%\bibitem{taji 112}\bysame, ``Simultaneous code/error-trellis reduction for convolutional codes using shifted code/error-subsequences,'' in {\em Proc. ISIT2011}, pp.~2409--2413, Aug. 2011.
\bibitem{taji 121}\bysame, ``Error-trellis construction for tailbiting convolutional codes,'' {\em IEICE Technical Report}, IT2011-59, pp.~85--90, March 2012.
\bibitem{taji 122}M.~Tajima and K.~Okino, ``Error-trellises for tailbiting convolutional codes,'' in {\em Proc. ISITA2012}, pp.~648--652, Oct. 2012.
\bibitem{taji 14}M.~Tajima, K.~Okino, and T.~Murayama, ``Tail-biting code/error-trellis construction for linear block codes,'' {\em IEICE Technical Report}, IT2014-39, pp.~153--158, July 2014.
\bibitem{taji 151}M.~Tajima, ``Tail-biting trellises for a tail-biting convolutional code,'' {\em IEICE Technical Report}, IT2015-25, pp.~47--52, July 2015.
\bibitem{taji 152}\bysame, ``Tail-biting trellis construction for linear block codes based on coset decomposition,'' in {\em Proc. SITA2015 (Japanese Ed.)}, pp.~680--685, Nov. 2015.
%\bibitem{tan 02}H.-H.~Tang and M.-C.~Lin, ``On (n, n-1) convolutional codes with low trellis complexity,'' {\em IEEE Trans. Commun.}, vol.~50, no.~1, pp.~37--47, Jan. 2002.
%\bibitem{tan 06}H.-H.~Tang, M.-C.~Lin, and B.~F.~U.~Filho, ``Minimal trellis modules and equivalent convolutional codes,'' {\em IEEE Trans.~Inform.~Theory}, vol.~52, no.~8, pp.~3738--3746, Aug. 2006.
%\bibitem{tann 04}R.~M.~Tanner, D.~Sridhara, A.~Sridharan, T.~E.~Fuja, and D.~J.~Costello, Jr., ``LDPC block and convolutional codes based on circulant matrices,'' {\em IEEE Trans.~Inform.~Theory}, vol.~50, no.~12, pp.~2966--2984, Dec. 2004.
\bibitem{tava 07}M.~B.S.~Tavares, K.~Sh.~Zigangirov, and G.~P.~Fettweis, ``Tail-biting LDPC convolutional codes,'' in {\em Proc. ISIT2007}, pp.~2341--2345, June 2007.
\bibitem{wea 12}E.~A.~Weaver, ``Minimality and duality of tail-biting trellises for linear codes,'' Ph.D. dissertation, submitted to the University of Kentucky, Lexington, Kentucky, USA, April 2012.
%\bibitem{wen 90}K.-A.~Wen, T.-S.~Wen, and J.-F.~Wang, ``A new transform algorithm for Viterbi decoding,'' {\em IEEE Trans.~Commun.}, vol.~38, no.~6, pp.~764--772, June 1990.
\bibitem{wo 78}J.~K.~Wolf, ``Efficient maximum likelihood decoding of linear block codes using a trellis,'' {\em IEEE Trans.~Inform.~Theory}, vol.~IT-24, no.~1, pp.~76--80, Jan. 1978.
%\bibitem{woz 57}J.~M.~Wozencraft, ``Sequential decoding for reliable communication,'' {\em MIT-RLE Tech. Rep.}, TR325, 1957.
\bibitem{yamada 83}T.~Yamada, H.~Harashima, and H.~Miyakawa, ``A new maximum likelihood decoding of high rate convolutional codes using a trellis,'' {\em Trans. IEICE (Japanese Ed.)}, vol.~J66-A, no.~7, pp.~611--616, July 1983.
\bibitem{zya 93}V.~Zyablov and V.~Sidorenko, ``Bounds on complexity of trellis decoding of linear block codes,'' {\em Probl. Pered. Inform.}, vol.~29, no.~3, pp.~3--9, July--Sept. 1993.

  %0.5em minus 0.4em\relax Harlow, England: Addison-Wesley, 1999.

\end{thebibliography}
